\documentclass[english]{IEEEtran}
\usepackage[T1]{fontenc}
\usepackage[latin9]{inputenc}
\usepackage{color}
\usepackage{float}
\usepackage{mathrsfs}
\usepackage{amsmath}
\usepackage{amssymb}
\usepackage{graphicx}
\usepackage{setspace}

\makeatletter

\providecommand{\tabularnewline}{\\}
\floatstyle{ruled}
\newfloat{algorithm}{tbp}{loa}
\providecommand{\algorithmname}{Algorithm}
\floatname{algorithm}{\protect\algorithmname}

\usepackage{cite}
\usepackage{times}
\usepackage{URL}

\usepackage{algorithmic}

\usepackage{subfig}

\makeatother

\usepackage{babel}
\begin{document}
\title{Robust Super-Resolution Compressive Sensing: A Two-timescale Alternating
MAP Approach}
\author{{\normalsize Yufan Zhou, Jingyi Li, Wenkang Xu, and An Liu, }~\IEEEmembership{Senior Member,~IEEE}{\normalsize\thanks{Yufan Zhou, Jingyi Li, Wenkang Xu, and An Liu are with the College
of Information Science and Electronic Engineering, Zhejiang University,
Hangzhou 310027, China (email: yufanzhou@zju.edu.cn, jingyili2003@zju.edu.cn,
22131113@zju.edu.cn, anliu@zju.edu.cn).}}}
\maketitle
\begin{abstract}
The problem of super-resolution compressive sensing (SR-CS) is crucial
for various wireless sensing and communication applications. Existing
methods often suffer from limited resolution capabilities and sensitivity
to hyper-parameters, hindering their ability to accurately recover
sparse signals when the grid parameters do not lie precisely on a
fixed grid and are close to each other. To overcome these limitations,
this paper introduces a novel robust super-resolution compressive
sensing algorithmic framework using a two-timescale alternating maximum
a posteriori (MAP) approach. At the slow timescale, the proposed framework
iterates between a sparse signal estimation module and a grid update
module. In the sparse signal estimation module, a hyperbolic-tangent
prior distribution based variational Bayesian inference (tanh-VBI)
algorithm with a strong sparsity promotion capability is adopted to
estimate the posterior probability of the sparse vector and accurately
identify active grid components carrying primary energy under a dense
grid. Subsequently, the grid update module utilizes the Broyden--Fletcher--Goldfarb--Shanno
(BFGS) algorithm to refine these low-dimensional active grid components
at a faster timescale to achieve super-resolution estimation of the
grid parameters with a low computational cost. The proposed scheme
is applied to the channel extrapolation problem, and simulation results
demonstrate the superiority of the proposed scheme compared to baseline
schemes.
\end{abstract}


\section{Introduction}

The problem of super-resolution compressive sensing (SR-CS) has attracted
a lot of research attention due to its wide applications in wireless
sensing and communication \cite{Fang_SR_CS,Yang_OGVBI}. The fundamental
goal of this problem is to recover a sparse signal $\boldsymbol{x}\in\mathbb{C}^{N\times1}$
from measurements $\boldsymbol{y}\in\mathbb{C}^{M\times1}$ ($M$
is often much less than $N$) under a linear observation model with
a dynamic grid:
\begin{equation}
\boldsymbol{\boldsymbol{y}}=\boldsymbol{A}\left(\boldsymbol{\theta}\right)\boldsymbol{x}+\boldsymbol{w},\label{eq:strandard model}
\end{equation}
where $\boldsymbol{A}\left(\boldsymbol{\theta}\right)\in\mathbb{C}^{M\times N}$
is the sensing matrix that depends on grid parameter $\boldsymbol{\theta}$,
and $\boldsymbol{w}\in\mathbb{C}^{M\times1}$ is the noise vector.
For instance, in narrowband integrated sensing and communication (ISAC)
systems \cite{Huang_ISAC,2022_ISAC_CE}, $\boldsymbol{x}$ can represent
both the angular-domain radar echo channel between the base station
(BS) and targets, as well as the angular-domain communication channel
between the user and the BS, while $\boldsymbol{\theta}$ corresponds
to the angle grid. Similarly, in multiple-input multiple-output orthogonal
frequency division multiplexing (MIMO-OFDM) channel estimation \cite{Kuai_MIMO_OFDM_CE,akbarpour_MIMO_OFDM_CE},
$\boldsymbol{x}$ denotes the angular-delay domain communication channel,
with $\boldsymbol{\theta}$ representing the angle-delay grid.

In conventional CS problems, $\boldsymbol{\theta}$ is typically fixed
and pre-determined. However, in many practical applications, the true
grid parameters $\boldsymbol{\theta}$ usually do not lie exactly
on the pre-determined fixed grid. If we use such a fixed grid, the
estimation accuracy of $\boldsymbol{\theta}$ will be limited by the
grid resolution, which will also cause energy leakage and degrade
performance. Moreover, it is important to achieve ``super-resolution''
estimation of the grid parameter in many scenarios, such as channel
extrapolation \cite{WAN_TWC} or high accuracy wireless sensing \cite{Xu_SLA_VBI},
especially if we want to separate multiple paths or targets that are
close to each other. In all such cases, dynamically adjusting the
grid is essential to achieve more accurate estimation of the grid
parameter $\boldsymbol{\theta}$ and alleviate the energy leakage
effect. Therefore, it is very important to consider the SR-CS problem.

Expectation-maximization (EM) framework based algorithms are recognized
as state-of-the-art (SOTA) methods for solving SR-CS problem \cite{SC_VBI,Liu_turbo_VBI}.
In this framework, the E-step employs sparse Bayesian learning (SBL)
algorithms to obtain the posterior probabilities of the sparse signal
and other latent variables, and the M-step updates the dynamic grid.
Iteration between two steps facilitates automatic learning of unknown
parameters, leading to superior performance and notable robustness
against uncertain parameters \cite{EM_algorithm}. For instance, the
authors in \cite{Liu_turbo_VBI} proposed a EM-based turbo variational
Bayesian inference (Turbo-VBI) algorithm for channel estimation in
massive MIMO systems.

Despite the demonstrated advantages of EM-based CS algorithms, existing
EM-based algorithms still have the following drawbacks. Firstly, the
existing SBL algorithms in the E-step model the sparsity using conditional
Gaussian \cite{Liu_turbo_VBI} or Laplace \cite{2008_TSP_laplace}
prior distributions, which roughly corresponds to the use of $l_{2}$-norm
or $l_{1}$-norm to model the number of non-zero elements in the sparse
signal. As a result, the sparsity promotion capability of these SBL
algorithms is typically limited, making it less robust under dense
grid when there are multiple close grid parameters, caused by e.g.,
multiple close paths or targets. Besides, conventional M-step often
adopts the gradient descent algorithm \cite{Liu_turbo_VBI,Dai_2018_fdd}
to update dynamic grid, which is susceptible to converging to local
optima when the objective functions are highly non-convex, thereby
often failing to achieve super-resolution grid update. Finally, EM-based
methods typically contain two-loop of iterations, namely, the inner
iterations in the E-step and the outer iteration between the E-step
and M-step, resulting in relatively slow convergence speed and high
computational complexity.

Recently, several research efforts have been dedicated to developing
more effective SR-CS algorithms. In \cite{WAN_TWC}, the authors formulated
channel extrapolation as a CS problem and then proposed a two-stage
channel extrapolation scheme. Firstly, the spatial and temporal multiple
signal classification (ST-MUSIC) algorithm is employed for initial
angle and delay estimation. As a subspace-based algorithm, ST-MUSIC
\cite{TST_MUSIC} is recognized for its inherent capability to achieve
high-resolution parameter estimation. Then an EM-based Turbo CS algorithm
is adopted for channel tracking, which is capable of exploiting the
temporal correlation of the channel and super-resolution prior information
of angle-delay parameters obtained from ST-MUSIC. Additionally, the
authors in \cite{DMRA} introduced the Dynamic Multi-Resolution of
Atoms (DMRA) algorithm, specifically tailored for dense line spectrum
super-resolution estimation. The DMRA algorithm utilizes a smooth
hyperbolic-tangent (tanh) relaxation function as an alternative to
$\ell_{0}\textrm{-norm}$ to effectively encourage sparsity, and performs
joint estimation of dominant grid components and their associated
complex gains. In a very recent work \cite{QNOMP}, the authors proposed
the Quasi-Newton Orthogonal Matching Pursuit (QNOMP) algorithm, which
is a two-stage super-resolution recovery process. It initiates with
an on-grid OMP estimation to identify dominant grid components, which
are subsequently refined through an off-grid optimization stage using
the Broyden--Fletcher--Goldfarb--Shanno (BFGS) method, thereby
enhancing both convergence speed and estimation accuracy.

Nevertheless, there are limitations of the existing works for solving
SR-CS problem:
\begin{itemize}
\item \textbf{Limited resolution capability}. Although the subspace-based
MUSIC algorithm can be adopted for initial estimation before constructing
the dynamic grid (sensing matrix) for SR-CS problem, its performance
is limited by certain operational conditions. Specifically, the MUSIC
algorithm typically requires a substantial number of measurements
(corresponding to the dimension of the observation vector) and multiple
independent snapshots to accurately construct the covariance matrix
\cite{TST_MUSIC}. In scenarios with limited measurement dimensions,
insufficient snapshots, or low signal-to-noise ratio (SNR) conditions,
the ability of MUSIC to resolve closely spaced paths degrades significantly.
Consequently, the MUSIC algorithm may yield inaccurate initial estimation,
which will severely restrict the performance of subsequent EM-based
CS algorithms, since EM-based algorithms often lack inherent super-resolution
capabilities and are highly dependent on the accuracy of initial parameter
estimation.
\item \textbf{Sensitivity to hyper-parameters and noise}. While algorithms
such as DMRA and QNOMP can demonstrate superior super-resolution capabilities
than traditional EM-based CS methods, their performance is sensitive
to hyper-parameter selection and noise. For the greedy-based QNOMP
algorithm \cite{QNOMP}, the on-grid OMP selection stage is sensitive
to noise, which can lead to incorrect index selection and error propagation
in subsequent steps. Furthermore, its stopping criterion relies on
the Constant False Alarm Rate (CFAR) principle, which is dependent
on an accurate noise variance estimate---a value that is often difficult
to obtain in practice. For the optimization-based DMRA algorithm \cite{DMRA},
while its source paper highlights its robustness within recommended
parameter ranges, achieving optimal performance still requires careful
tuning, as the ideal parameter settings may vary across different
application scenarios. Improper selection of these hyper-parameters
can lead to a substantial performance degradation, thereby limiting
the practical deployment of these methods.
\item \textbf{Lack of a unified algorithmic framework and limited applicability
to general cases.} A critical limitation of these advanced methods
such as DMRA and QNOMP is the lack of a unified algorithmic framework,
which results in limited applicability to general cases. Specifically,
the DMRA algorithm is composed of a multi-stage pipeline, and the
entire workflow is fundamentally based on the properties of the Discrete
Fourier Transform (DFT). Consequently, the DMRA algorithm is incompatible
with general compressed sensing problems involving arbitrary sensing
matrices. For instance, when applied to wireless communication, this
structural limitation means the algorithm is well-suited for Uniform
Linear Arrays (ULA) but not directly applicable to more complex antenna
architectures like Uniform Planar Arrays (UPA) without substantial
modifications.
\end{itemize}

Motivated by the limitations of existing algorithms, this paper introduces
a novel solution for the SR-CS problem, which is based on a two-timescale
alternating maximum a posteriori (MAP) framework. The main contributions
of this paper are summarized as follows.
\begin{itemize}
\item \textbf{A two-timescale alternating MAP framework:} We propose a novel
two-timescale alternating MAP framework specifically tailored for
the SR-CS problem. This framework operates on a two-timescale fashion.
At the slow timescale, it contains two key modules: the sparse signal
estimation module and the grid update module. The sparse signal estimation
module estimates the posterior probability of the sparse vector, effectively
identifying potential active grid components that carry the primary
energy under a dense grid. Subsequently, the grid update module refines
both the active grid components and their complex gains at a faster
timescale, due to the following considerations: 1) more frequent update
of the active grid leads to higher precision of grid refinement and
stronger resolution capability and 2) the number of active grid components
is usually small in CS and thus the computational cost of updating
them is relatively small. As such, the two modules work iteratively
to jointly estimate the sparse vector and update grid parameters at
different timescales, enhancing the resolution capability at minimum
computational cost.
\item \textbf{Tanh-VBI algorithm for sparse signal estimation module:} The
sparse signal estimation module needs to accommodate dense grid caused
by multiple close paths. However, conventional SBL algorithms used
in the EM-based CS algorithms or the on-grid OMP used in the QNMOP
usually cannot work well under a dense grid. To address this challenge,
we introduce a novel tanh distribution based variational Bayesian
inference (tanh-VBI) algorithm to estimate the posterior probability
of the sparse signal under a dense grid. Specially, the tanh-VBI algorithm
uses a conditional tanh distribution (based on a better approximation
of the $l_{0}$-norm) to model the prior of sparse signals, which
offers a stronger sparsity promotion capability compared to conventional
approaches that use Gaussian (based on $l_{2}$-norm) or Laplace (based
on $l_{1}$-norm) prior distributions\footnote{It is observed that the resolution capability is closely related to
the sparsity promotion capability.}. By employing a successive linear approximation approach, the variational
Bayesian inference can be performed in a closed form.
\item \textbf{BFGS algorithm for grid update module:} For given active grid
components and corresponding estimated posterior probabilities from
the sparse signal estimation module, we adopt the BFGS algorithm for
adjusting grid parameters. As a quasi-Newton method, BFGS approximates
the inverse Hessian matrix, thereby incorporating second-order derivative
information of the posterior function with respect to the grid parameters.
This enables BFGS to determine more effective search direction and
step-size selection compared to methods relying solely on first-order
gradient information, such as traditional gradient descent. Consequently,
the BFGS algorithm typically exhibits greater robustness and is less
prone to becoming trapped in local optimal, particularly in highly
non-convex optimization problems. Moreover, the low-dimensional active
grid parameters and their complex gains are updated at a faster timescale
to accelerate the convergence and improve the grid resolution. As
such, the fast-timescale BFGS algorithm facilitates more accurate
grid refinement than traditional gradient descent methods, which is
critical for achieving super-resolution estimation.
\end{itemize}

The remainder of this paper is organized as follows. In Section \ref{sec:System-Model},
we introduce a three-layer sparse prior model with stronger sparsity
promotion capability, formulate the SR-CS problem and discuss its
practical application in wireless communication. In Section \ref{sec:The-proposed-Alternating},
we introduce the proposed two-timescale alternating MAP framework,
providing an overview of its structure and detailing the grid update
module. The sparse signal estimation module, which leverages the tanh-VBI
algorithm, is presented in Section \ref{sec:Turbo-VBI-Algorithm}.
Simulations applied to a channel extrapolation problem are shown in
Section \ref{sec:Simulations}. Finally, the conclusion is given in
Section \ref{sec:Conlusion}.

\textit{Notation:} Lowercase boldface letters denote vectors and uppercase
boldface letters denote matrices. $\left(\cdot\right)^{-1}$, $\left(\cdot\right)^{T}$,
$\left(\cdot\right)^{H}$, $\mid\cdot\mid$, $\left\Vert \cdot\right\Vert $,
and $\left\langle \cdot\right\rangle $ are used to represent the
inverse, transpose, conjugate transpose, magnitude, $\ell_{2}\textrm{-norm}$,
and expectation operations, respectively. $\mathbf{I}_{M}$ denotes
the $M\times M$ dimensional identity matrix. For a vector $\boldsymbol{x}\in\mathbb{C}^{N}$
and a given index set $\mathcal{S}\subseteq\left\{ 1,...,N\right\} $,
$\left|\mathcal{S}\right|$ denotes its cardinality, $\boldsymbol{x}_{\mathcal{S}}\in\mathbb{C}^{\left|\mathcal{S}\right|\times1}$
denotes the subvector consisting of the elements of $\boldsymbol{x}$
indexed by the set $\mathcal{S}$. $\text{diag}\left(\boldsymbol{x}\right)$
denotes a block diagonal matrix with $\boldsymbol{x}$ as the diagonal
elements. $\mathcal{CN}\left(\boldsymbol{x};\boldsymbol{\mu},\mathbf{\Sigma}\right)$
represents a complex Gaussian distribution with mean $\boldsymbol{\mu}$
and covariance matrix $\mathbf{\Sigma}$. $\Gamma\left(x;a,b\right)$
represents a Gamma distribution with shape parameter $a$ and rate
parameter $b$. Finally, $x=\Theta(a)$ for $a>0$ denotes that $\exists k_{1},k_{2}>0$,
such that $k_{2}\cdot a\leq x\leq k_{1}\cdot a$.

\section{SR-CS Problem Formulation\label{sec:System-Model}}

\subsection{Observation Model in SR-CS}

Recall that in SR-CS problem, the observation model can be written
in a linear form:

\begin{equation}
\boldsymbol{y}=\mathbf{A}\left(\boldsymbol{\theta}\right)\boldsymbol{x}+\boldsymbol{w},\label{eq:general_model}
\end{equation}
where $\boldsymbol{x}\in\mathbb{C}^{N\times1}$ is the sparse signal,
$\boldsymbol{y}\in\mathbb{C}^{M\times1}$ is the observation vector,
$\mathbf{A}\left(\boldsymbol{\theta}\right)\in\mathbb{C}^{M\times N}$
is the sensing matrix that depends on grid parameter $\boldsymbol{\theta}\in\mathbb{C}^{N\times1}$,
and $\boldsymbol{w}\in\mathbb{C}^{M\times1}$ is the noise vector
with independent Gaussian entries $w_{m}\sim\mathcal{CN}\left(w_{m};0,\kappa^{-1}\right)$. 

We employ a Gamma distribution with parameters $c$ and $d$ to model
the noise precision, i.e.,
\begin{equation}
p\left(\kappa\right)=\Gamma\left(\kappa;c,d\right).\label{eq:p(gamma)}
\end{equation}
The Gamma distribution can capture the practical distribution of the
noise precision well and is a conjugate of the Gaussian prior. Therefore,
it has been widely used to model the noise precision in Bayesian inference
\cite{Tzikas_VBI,Liu_turbo_VBI,Xu_SLA_VBI}. 

Note that in conventional CS problems, the grid parameter $\boldsymbol{\theta}$
is typically initialized as a uniform grid, and the interval between
adjacent grid points is constrained by the system's configuration.
For instance, in wireless communication systems, the interval of a
uniform delay grid is inversely proportional to the system's bandwidth.
However, in SR-CS problems, a grid denser than the above uniform grid
should be introduced to resolve closely-spaced paths and achieve super-resolution
estimation, which poses new challenges in CS algorithm design since
a denser grid leads to higher correlation between the columns of the
sensing matrix $\mathbf{A}\left(\boldsymbol{\theta}\right)$. Conventional
CS algorithms such as OMP or SBL may not work well under dense grid
and we have to design more powerful CS algorithms with stronger sparsity
promotion capability.

\subsection{Three-layer Bernoulli-Gamma-Tanh Sparse Prior Model}

The sparse prior probability model forms the foundation for Bayesian
inference in sparse signal recovery. In \cite{Liu_turbo_VBI}, the
authors introduced a three-layer sparse prior model, which is flexible
to capture various sparse structures and robust to imperfect prior
information in practice. However, the sparsity promotion capability
of this model is still limited and may not works well under dense
grid, as it models sparsity using conditional Gaussian priors, which
effectively corresponds to the use of the $l_{2}$-norm. 

\begin{figure}
\centering{}\includegraphics[width=75mm]{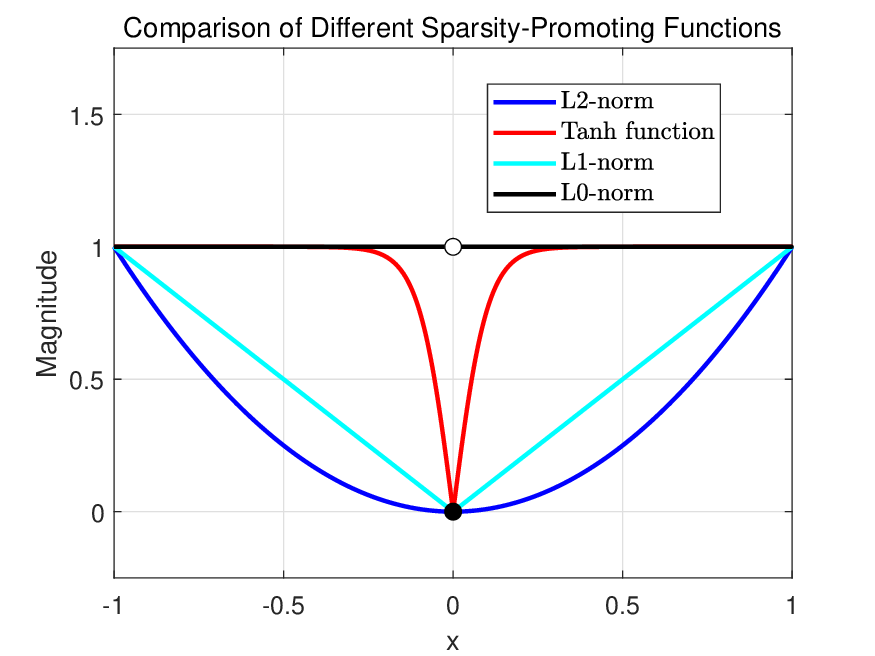}\caption{\label{fig:relax_function}An illustration of different relaxation
functions.}
\end{figure}

Generally, the closer a smooth relaxation is to the ideal $l_{0}$-norm,
the stronger its ability to promote sparsity. As discussed in \cite{DMRA},
the tanh function, i.e, $\textrm{tanh}\left(x\right)=\frac{\textrm{e}^{x}-\textrm{e}^{-x}}{\textrm{e}^{x}+\textrm{e}^{-x}}$,
provides a more accurate relaxation to the ideal $l_{0}$-norm than
both the $l_{2}$-norm and $l_{1}$-norm, as illustrated in Fig. \ref{fig:relax_function}.
Specifically, the tanh function makes small coefficients much more
likely to be zero, while for larger values the penalty remains almost
unchanged. This results in a clearer separation between zero and nonzero
components, similar to the effect of the ideal $l_{0}$-norm.

Motivated by this intuition, we propose a novel three-layer Bernoulli-Gamma-Tanh
(BGT) sparse prior model that utilizes the tanh function to model
the conditional prior distribution of the sparse signal $\boldsymbol{x}$,
referred to as the tanh distribution. As expected, compared with commonly
used Gaussian and Laplace priors, the tanh distribution based prior
can better approximate the ideal $l_{0}$-norm, providing a sharper
distinction between zero and nonzero coefficients. Consequently, the
tanh distribution based prior is expected to achieve stronger sparsity
promotion capability, making it particularly suitable for the SR-CS
problems with dense grid.

\begin{figure}
\centering{}\includegraphics[width=75mm]{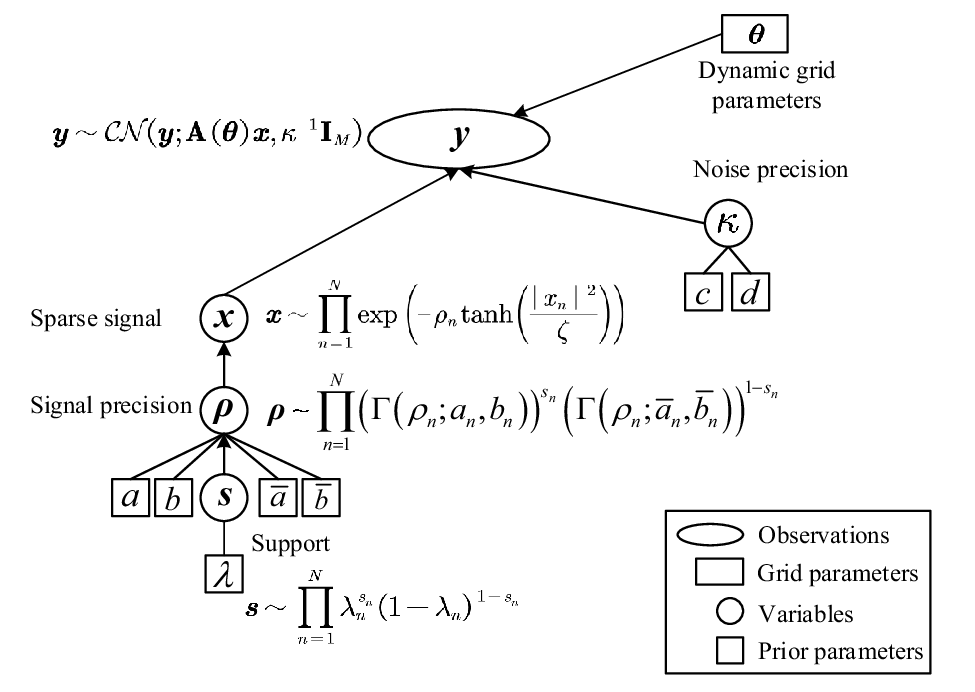}\caption{\label{fig:3LHS-Model}An illustration of three-layer hierarchical
sparse prior model.}
\end{figure}

In the proposed three-layer BGT sparse prior, a support vector $\boldsymbol{s}\triangleq\left[s_{1},\ldots,s_{N}\right]^{T}\in\left\{ 0,1\right\} ^{N}$
is introduced to indicate whether the $n$-th element $x_{n}$ in
$\boldsymbol{x}$ is active ($s_{n}=1$) or inactive ($s_{n}=0$).
Specifically, let $\boldsymbol{\rho}=\left[\rho_{1},...,\rho_{N}\right]^{T}$
denote the precision vector of $\boldsymbol{x}$ (i.e., $1/\rho_{n}$
denotes the variance of $x_{n}$). Then the joint distribution of
$\boldsymbol{x}$, $\boldsymbol{\rho}$, and $\boldsymbol{s}$ can
be expressed as
\begin{equation}
p\left(\boldsymbol{x},\boldsymbol{\rho},\boldsymbol{s}\right)=\underbrace{p\left(\boldsymbol{s}\right)}_{\textrm{Support}}\underbrace{p\left(\boldsymbol{\rho}\mid\boldsymbol{s}\right)}_{\textrm{Precision}}\underbrace{p\left(\boldsymbol{x}\mid\boldsymbol{\rho}\right)}_{\textrm{Sparse\ signal}},\label{eq:p(x,rou,s)}
\end{equation}
as illustrated in Fig. \ref{fig:3LHS-Model}. In the following, we
detail the probability model for each variable.

The prior distribution $p\left(\boldsymbol{s}\right)$ of the support
vector is used to capture the sparsity in specific applications. For
example, to capture an independent sparse structure, we can set 
\begin{equation}
p\left(\boldsymbol{s}\right)=\prod_{n=1}^{N}\left(\lambda_{n}\right)^{s_{n}}\left(1-\lambda_{n}\right)^{1-s_{n}},\label{eq:iid_support}
\end{equation}
where $\lambda_{n}$ is the spar\textcolor{black}{sity ratio. Note
that for clarity, this paper focuses on the independent sparse prior
in \eqref{eq:iid_support}. However, our proposed algorithm also works
for a general choice of }$p\left(\boldsymbol{s}\right)$ by applying
the Turbo approach to combining a structured sparse inference module
(to handle more complicated structured sparse prior $p\left(\boldsymbol{s}\right)$
via message passing) and the Tanh-VBI algorithm\textcolor{black}{,}
as proposed in \textcolor{black}{\cite{Liu_turbo_VBI}.}
\begin{figure}
\centering{}\includegraphics[width=75mm]{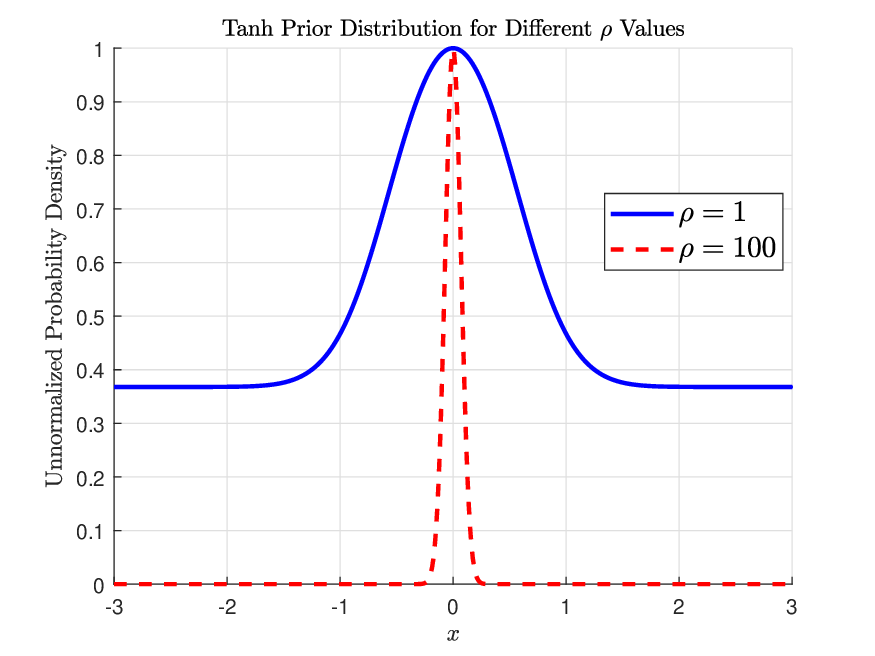}\caption{\label{fig:show_different_rou}An illustration of tanh distribution
with different $\rho$ values.}
\end{figure}

The conditional probability $p\left(\boldsymbol{\rho}\mid\boldsymbol{s}\right)$
is given by
\begin{align}
p\left(\boldsymbol{\rho}\mid\boldsymbol{s}\right) & =\prod_{n=1}^{N}\left(\Gamma\left(\rho_{n};a_{n},b_{n}\right)\right)^{s_{n}}\left(\Gamma\left(\rho_{n};\overline{a}_{n},\overline{b}_{n}\right)\right)^{1-s_{n}},\label{eq:rou_cond_s}
\end{align}
where $\Gamma\left(\rho;a,b\right)$ is a Gamma hyper-prior with shape
parameter $a$ and rate parameter $b$. When $s_{n}=1$, the variance
$1/\rho_{n}$ of $x_{n}$ is $\Theta\left(1\right)$, and thus the
shape and rate parameters $a_{n},b_{n}$ should be chosen such that
$\frac{a_{n}}{b_{n}}=\mathbb{E}\left[\rho_{n}\right]=\Theta\left(1\right)$.
On the other hand, when $s_{n}=0$, $x_{n}$ is close to zero, and
thus the shape and rate parameters $\overline{a}_{n},\overline{b}_{n}$
should be chosen to satisfy $\frac{\overline{a}_{n}}{\overline{b}_{n}}=\mathbb{E}\left[\rho_{n}\right]\gg1$.

The conditional probability $p\left(\boldsymbol{x}\mid\boldsymbol{\rho}\right)$
for the sparse signal is assumed to have a product form $p\left(\boldsymbol{x}\mid\boldsymbol{\rho}\right)=\prod_{n=1}^{N}p\left(x_{n}\mid\rho_{n}\right)$
and each $p\left(x_{n}\mid\rho_{n}\right)$ is modeled as a tanh distribution
to achieve better sparsity promotion capability:
\begin{equation}
p\left(x_{n}\mid\rho_{n}\right)=\frac{1}{C\left(\rho_{n},\zeta\right)}\textrm{exp}\left(-\rho_{n}\textrm{tanh}\left(\frac{\mid x_{n}\mid^{2}}{\zeta}\right)\right),\label{eq:element_wise_tanh}
\end{equation}
where $\zeta$ is the relaxation parameter, and $C\left(\rho_{n},\zeta\right)=\int\textrm{exp}\left(-\rho_{n}\textrm{tanh}\left(\frac{\mid x_{n}\mid^{2}}{\zeta}\right)\right)\textrm{d}x_{n}$
is the normalized constant with respect to $x_{n}$. Note that when
$\mathbb{E}\left[\rho_{n}\right]\gg1$, the distribution (\ref{eq:element_wise_tanh})
becomes extremely peaked at zero, which strongly promotes sparsity,
as illustrated in Fig. \ref{fig:show_different_rou}. In practice,
$\zeta$ is typically set within the range of $\left[0,1\right]$,
where a smaller value corresponds to stronger sparsity promotion.
A common strategy is to adapt $\zeta$ according to the SNR, employing
a smaller $\zeta$ for higher SNR conditions. Following this guidance,
the performance of the proposed algorithm is not highly sensitive
to the precise value of $\zeta$ within a reasonable range. 

Note that when $x_{n}\rightarrow\infty$, $\textrm{exp}\left(-\rho_{n}\textrm{tanh}\left(\frac{\mid x_{n}\mid^{2}}{\zeta}\right)\right)$
trends to $\textrm{exp}\left(-\rho_{n}\right)\neq0$. As a result,
integrating over the entire domain of $x_{n}$ would lead to divergence.
To ensure that the normalization constant $C\left(\rho_{n},\zeta\right)$
remains finite, we restrict the domain of the variable, i.e, $\mid x_{n}\mid\leq X_{\textrm{max}},\forall n$.
With this constraint, $C\left(\rho_{n},\zeta\right)$ is guaranteed
to converge.

\subsection{Problem Formulation \label{sec:CS-Problem-Formulation}}

Given the observation $\boldsymbol{y}$, we aim at computing a Bayesian
estimation of the sparse signal $\boldsymbol{x}$ and support $\boldsymbol{s}$,
i.e., the posterior $p\left(\boldsymbol{x}\mid\boldsymbol{y}\right)$
and $p\left(\boldsymbol{s}\mid\boldsymbol{y}\right)$, and the maximum
likelihood estimation (MLE) of grid parameters $\boldsymbol{\theta}$,
i.e., $\underset{\boldsymbol{\theta}}{\text{argmax}}\ln p\left(\boldsymbol{\theta}\mid\boldsymbol{y}\right)$\footnote{Note that the MAP estimator reduces to the MLE since we assume no
prior knowledge on the grid parameters $\boldsymbol{\theta}$.}. Note that we can obtain an MAP estimate of $\boldsymbol{x}$ and
$\boldsymbol{s}$ from the Bayesian estimation $p\left(\boldsymbol{x}\mid\boldsymbol{y}\right)$
and $p\left(\boldsymbol{s}\mid\boldsymbol{y}\right)$. In the following,
we give a concrete application example of the above SR-CS problem.

\subsection{Practical Application: Channel Extrapolation\label{subsec:Example:-Massive-MIMO}}

\begin{figure}
\begin{centering}
\includegraphics[width=85mm]{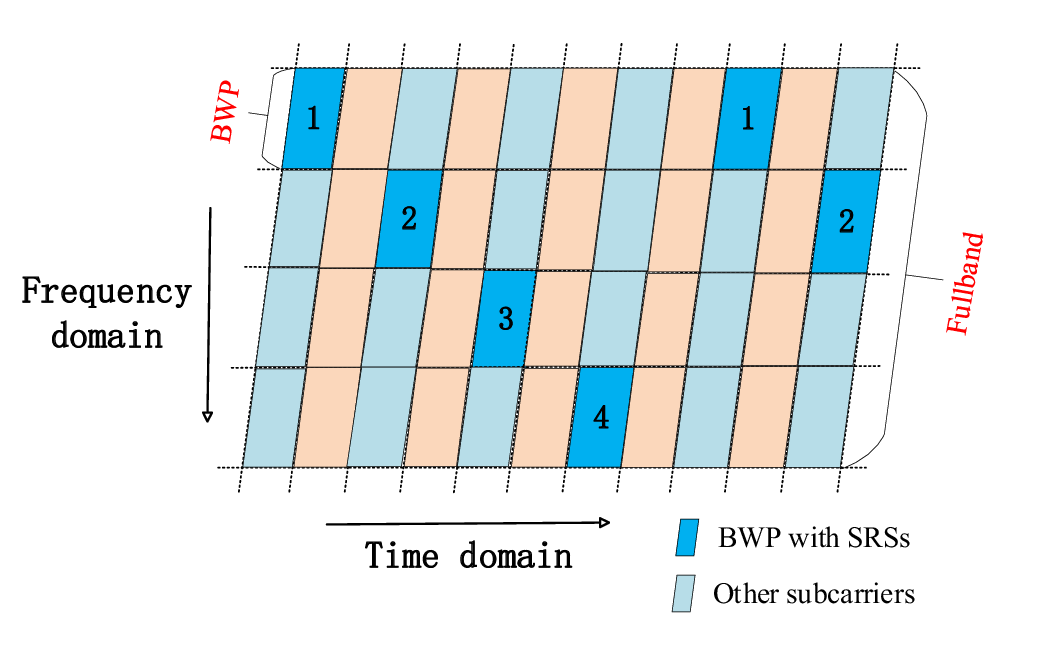}
\par\end{centering}
\caption{\label{fig:Illustration_BS}An illustration of channel extrapolation
for $h_{p}=4$.}
\end{figure}

In practical time division duplex (TDD) massive multiple-input multiple-output
orthogonal frequency-division multiplexing (MIMO-OFDM) systems, channel
extrapolation is an essential technique for efficient channel state
information (CSI) acquisition. Due to limited transmission power,
the user typically transmits uplink pilot signals only within a Bandwidth
Part (BWP), occupying only a fraction of the total system bandwidth
\cite{3gpp_5GNR,WAN_TWC}, as illustrated in Fig. \ref{fig:Illustration_BS}.
This constraint makes it necessary to estimate the fullband channel
based on the measurements with limited bandwidth.

Consider a typical TDD massive MIMO-OFDM system, where a base station
(BS) equipped with $N_{r}$ uniform linear array (ULA) antennas serves
a single-antenna user. The entire system bandwidth is partitioned
into $h_{p}$ BWPs, each containing $M$ subcarriers, such that there
are $h_{p}M$ subcarriers in total. The channel frequency response
(CFR) at the $n$-th subcarrier ($0\leq n\leq h_{p}M-1$) can be expressed
as:
\begin{equation}
\boldsymbol{h}_{n}=\sum_{k=1}^{K}\alpha_{k}e^{-j2\pi nf_{0}\tau_{k}}\boldsymbol{a}_{R}\left(\theta_{k}\right),
\end{equation}
where $K$ is the number of propagation paths, $\alpha_{k}$ is the
complex gain of the $k\textrm{-th}$ path, $\tau_{k}$ is the delay
of the $k\textrm{-th}$ path, and $\boldsymbol{a}_{R}\left(\theta_{k}\right)\in\mathbb{C}^{N_{r}\times1}$
is the array response vector at angle $\theta_{k}$.

The received signal $\mathbf{Y}\in\mathbb{C}^{M\times N_{r}}$ at
the BS can be written as:
\begin{equation}
\mathbf{Y}=\textrm{diag}\left(\boldsymbol{\beta}\right)\mathbf{W}\mathbf{H}+\mathbf{N},\label{eq:received signal model}
\end{equation}
where $\boldsymbol{\beta}\in\mathbb{C}^{M\times1}$ denotes the uplink
pilot vector transmitted from user, $\mathbf{W}\in\{0,1\}^{M\times h_{p}M}$
is the subcarrier selection matrix, $\mathbf{H}=\left[\boldsymbol{h}_{0}^{T};\boldsymbol{h}_{1}^{T};\ldots;\boldsymbol{h}_{h_{p}M-1}^{T}\right]\in\mathbb{C}^{h_{p}M\times N_{r}}$
denotes the fullband CFR matrix, and $\mathbf{N}$ is additive white
Gaussian noise with each element having zero mean and variance $\sigma_{e}^{2}$.

To facilitate super-resolution estimation, a grid-based sparse representation
is employed. Specifically, a dense two-dimensional dynamic grid is
employed in the angular-delay domain. After obtaining coarse estimation
of the angle and delay parameters using low-complexity baseline algorithms,
e.g, the ST-MUSIC algorithm \cite{TST_MUSIC}, densely sampled grid
points are placed in the surrounding regions to better resolve closely
spaced path components.

Let $\left\{ \left(\theta_{q},\tau_{q}\right)\right\} _{q=1}^{Q}$
denotes the collection of grid points in the angular-delay domain,
where $Q$ is the total number. For convenience, we define the grid
vectors $\boldsymbol{\theta}\triangleq\left[\theta_{1},\ldots,\theta_{Q}\right]^{T}$
and $\boldsymbol{\tau}\triangleq\left[\tau_{1},\ldots,\tau_{Q}\right]^{T}$.
The received signal model in (\ref{eq:received signal model}) can
be reformulated as:
\begin{equation}
\ensuremath{\mathbf{Y}=\textrm{diag}\left(\boldsymbol{\beta}\right)\mathbf{W}\mathbf{B}\left(\boldsymbol{\tau}\right)\textrm{diag}\left(\boldsymbol{x}\right)\mathbf{A}\left(\boldsymbol{\theta}\right)^{T}+}\mathbf{N},
\end{equation}
with two dictionary matrices:
\begin{equation}
\mathbf{B}\left(\boldsymbol{\tau}\right)\triangleq\left[\boldsymbol{b}\left(\tau_{1}\right),\ldots,\boldsymbol{b}\left(\tau_{Q}\right)\right]\in\mathbb{C}^{h_{p}M\times Q},
\end{equation}
\begin{equation}
\mathbf{A}\left(\boldsymbol{\theta}\right)\triangleq\left[\boldsymbol{a}_{R}\left(\theta_{1}\right),\ldots,\boldsymbol{a}_{R}\left(\theta_{Q}\right)\right]\in\mathbb{C}^{N_{r}\times Q},
\end{equation}
where $\boldsymbol{b}\left(\tau_{q}\right)=\left[1,e^{-j2\pi f_{0}\tau_{q}},\ldots,e^{-j2\pi\left(h_{p}M-1\right)f_{0}\tau_{q}}\right]^{T}\in\mathbb{C}^{h_{p}M\times1}$,
and $\boldsymbol{x}\in\mathbb{C}^{Q\times1}$ is the angular-delay
domain sparse vector, which has only $K\ll Q$ non-zero elements corresponding
to $K$ paths. Specifically, the $q$-th element of $\boldsymbol{x}$,
denoted by $x_{q}$, is the complex gain of the channel path lying
around the $q$-th grid point.

To further facilitate algorithmic processing, the model can be vectorized
as:
\begin{equation}
\ensuremath{\boldsymbol{y}=\mathbf{\Phi}\left(\boldsymbol{\theta},\boldsymbol{\tau}\right)\boldsymbol{x}+}\boldsymbol{n},
\end{equation}
where $\boldsymbol{y}=\textrm{vec}\left(\mathbf{Y}\right)$, $\boldsymbol{n}=\textrm{vec}\left(\mathbf{N}\right)$,
and $\mathbf{\Phi}\left(\boldsymbol{\theta},\boldsymbol{\tau}\right)\in\mathbb{C}^{MN_{r}\times Q}$
is the sensing matrix with dynamic angle-delay grid, with its $q$-th
column given by $\boldsymbol{a}_{R}\left(\theta_{q}\right)\otimes\left[\textrm{diag}\left(\boldsymbol{\beta}\right)\mathbf{W}\boldsymbol{b}\left(\tau_{q}\right)\right]$,
which shares the same form as the general linear observation model
with unknown parameters in (\ref{eq:general_model}). For simplicity,
we focus on this general observation model in (\ref{eq:general_model})
for subsequent algorithm design and analysis throughout the remainder
of this paper.

\section{The Proposed Two-timescale Alternating MAP Framework\label{sec:The-proposed-Alternating}}

\subsection{Outline of the Proposed Algorithm \label{subsec:The-Grid-Estimation-1}}

It is very challenging to calculate the exact posterior $p\left(\boldsymbol{x}\mid\boldsymbol{y}\right)$,
$p\left(\boldsymbol{s}\mid\boldsymbol{y}\right)$ and the MLE solution
$\hat{\boldsymbol{\theta}}=\underset{\boldsymbol{\theta}}{\text{argmax}}\ln p\left(\boldsymbol{\theta}\mid\boldsymbol{y}\right)$,
because the factor graph of the associated joint probability model
has loops and the likelihood function is highly non-convex. To solve
this challenge, we propose a two-timescale alternating MAP framework
to approximately calculate the marginal posteriors $p\left(x_{n}\mid\boldsymbol{y}\right)$
and $p\left(s_{n}\mid\boldsymbol{y}\right),\forall n$, and finds
an approximate solution for MLE problem $\underset{\boldsymbol{\theta}}{\text{argmax}}\ln p\left(\boldsymbol{\theta}\mid\boldsymbol{y}\right)$.

\begin{figure}[t]
\begin{centering}
\includegraphics[width=90mm]{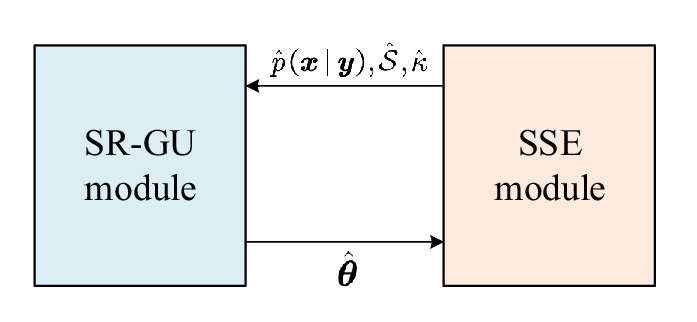}
\par\end{centering}
\caption{\label{fig:AE}The basic modules of the alternating estimation framework.}
\end{figure}

As illustrated in Fig. \ref{fig:AE}, the proposed framework alternates
between the following two modules until convergence, operating on
two different timescales.
\begin{itemize}
\item \textbf{Sparse signal estimation (SSE) module in slow timescale: }For
a fixed ML estimator $\hat{\boldsymbol{\theta}}$ of $\boldsymbol{\theta}$
obtained from the super-resolution grid update (SR-GU) module, the
SSE module leverages a tanh-VBI algorithm to perform variational Bayesian
inference for Bayesian estimation of the collection of parameters
$\boldsymbol{v}=\left\{ \boldsymbol{x},\boldsymbol{\rho},\boldsymbol{s},\kappa\right\} $.
This process outputs the estimated posterior distributions $\hat{p}\left(\boldsymbol{x}\mid\boldsymbol{y}\right)$
and $\hat{p}\left(\kappa\mid\boldsymbol{y}\right)$, which in turn
yield the MAP estimators: $\hat{\boldsymbol{x}}$ for the sparse signal,
$\hat{\kappa}$ for the noise precision $\kappa$. Finally, the estimated
support $\hat{\mathcal{S}}$ can be calculated from $\hat{\boldsymbol{x}}$.
\item \textbf{Super-resolution grid update module in fast timescale:} For
fixed $\hat{p}\left(\boldsymbol{x}\mid\boldsymbol{y}\right)$, $\hat{\mathcal{S}}$,
and $\hat{\kappa}$ output from the SSE module, the SR-GU module alternately
refines the active dynamic grid $\boldsymbol{\theta}_{\hat{\mathcal{S}}}$
and its corresponding complex gain $\hat{\boldsymbol{x}}_{\hat{\mathcal{S}}}$
on a faster timescale by MAP approach. In the grid update process,
the BFGS algorithm is employed to efficiently maximize the highly
non-convex likelihood function. Moreover, the step-size is carefully
chosen using the Armijo rule to ensure effective search along the
descent direction.
\end{itemize}

In the following, we present the details of the SR-GU module. The
proposed SSE module and associated tanh-VBI algorithm are described
in Section \ref{sec:Turbo-VBI-Algorithm}.

\subsection{The Super-Resolution Grid Update Module in Fast Timescale\label{subsec:The-Grid-Estimation}}

\subsubsection{MAP problem formulation}

To facilitate a more frequent and efficient update of the grid parameters
in the fast-timescale SR-GU module, we only retain the sparse signals
and grid parameters indexed by the estimated support $\hat{\mathcal{S}}$,
discarding all other signals and grid parameters. The variables $\boldsymbol{x}_{\hat{\mathcal{S}}}$
and $\boldsymbol{\theta}_{\hat{\mathcal{S}}}$ are updated in an alternating
manner. 

For a fixed $\boldsymbol{\theta}_{\hat{\mathcal{S}}}$ and $\hat{\kappa}$,
the signal vector $\boldsymbol{x}_{\hat{\mathcal{S}}}$ is updated
by solving a MAP problem. We leverage $\hat{p}\left(\boldsymbol{x}\mid\boldsymbol{y}\right)$
and $\hat{\mathcal{S}}$ obtained from the SSE module to construct
the prior distribution of $\boldsymbol{x}_{\hat{\mathcal{S}}}$. Specially,
the prior distribution of $\boldsymbol{x}_{\hat{\mathcal{S}}}$ is
modeled as a complex Gaussian distribution $\mathcal{CN}\left(\boldsymbol{x}_{\hat{\mathcal{S}}};\boldsymbol{u}_{\hat{\mathcal{S}}},\mathbf{\Sigma}_{\hat{\mathcal{S}}}\right)$,
where $\boldsymbol{u}_{\hat{\mathcal{S}}}$ is the mean of $\hat{p}\left(\boldsymbol{x}\mid\boldsymbol{y}\right)$,
and $\mathbf{\Sigma}_{\hat{\mathcal{S}}}$ is set to slightly larger
than the variance of $\hat{p}\left(\boldsymbol{x}\mid\boldsymbol{y}\right)$
to improve the robustness against the estimation error of the SSE
module. Consequently, the resulting MAP optimization problem of $\boldsymbol{x}_{\hat{\mathcal{S}}}$,
which is equivalent to a linear minimum mean squared error (LMMSE)
estimation problem \cite{estimation_theory_book}, is expressed as:
\begin{align}
\boldsymbol{\bar{x}}_{\hat{\mathcal{S}}}= & \arg\min_{\boldsymbol{x}_{\hat{\mathcal{S}}}}\ \mathcal{\psi}\left(\boldsymbol{x}_{\hat{\mathcal{S}}}\right)=\hat{\kappa}\left\Vert \boldsymbol{y}-\mathbf{A}_{\hat{\mathcal{S}}}\left(\boldsymbol{\theta}_{\hat{\mathcal{S}}}\right)\boldsymbol{x}_{\hat{\mathcal{S}}}\right\Vert ^{2}\nonumber \\
+ & \left(\boldsymbol{x}_{\hat{\mathcal{S}}}-\boldsymbol{u}_{\hat{\mathcal{S}}}\right)^{H}\mathbf{\Sigma}_{\hat{\mathcal{S}}}^{-1}\left(\boldsymbol{x}_{\hat{\mathcal{S}}}-\boldsymbol{u}_{\hat{\mathcal{S}}}\right),\label{eq:x_fun}
\end{align}
where $\mathbf{A}_{\hat{\mathcal{S}}}\left(\boldsymbol{\theta}_{\hat{\mathcal{S}}}\right)\in\mathbb{C}^{M\times\left|\hat{\mathcal{S}}\right|}$
is a sub-matrix of $\mathbf{A}$ with the column indices lying in
$\hat{\mathcal{S}}$. By setting the gradient of the objective function
$\mathcal{\psi}\left(\boldsymbol{x}_{\hat{\mathcal{S}}}\right)$ to
zero, we get the following closed-form solution of $\boldsymbol{x}_{\hat{\mathcal{S}}}$.
For notation simplicity, the dependency of $\mathbf{A}_{\hat{\mathcal{S}}}$
on $\boldsymbol{\theta}_{\hat{\mathcal{S}}}$ is omitted here:
\begin{equation}
\boldsymbol{\bar{x}}_{\hat{\mathcal{S}}}=\left(\mathbf{A}_{\hat{\mathcal{S}}}^{H}\mathbf{A}_{\hat{\mathcal{S}}}+\frac{1}{\hat{\kappa}}\mathbf{\Sigma}_{\hat{\mathcal{S}}}^{-1}\right)^{-1}\left(\mathbf{A}_{\hat{\mathcal{S}}}^{H}\boldsymbol{y}+\frac{1}{\hat{\kappa}}\mathbf{\Sigma}_{\hat{\mathcal{S}}}^{-1}\boldsymbol{u}_{\hat{\mathcal{S}}}\right),\label{LMMSE}
\end{equation}

Similarly, for a fixed $\boldsymbol{x}_{\hat{\mathcal{S}}}$, the
grid parameters $\boldsymbol{\theta}_{\hat{\mathcal{S}}}$ are updated
based on MLE. The MLE problem is formulated as:
\begin{equation}
\boldsymbol{\bar{\theta}}_{\hat{\mathcal{S}}}=\arg\min_{\boldsymbol{\theta}_{\hat{\mathcal{S}}}}\ \mathcal{L}\left(\boldsymbol{\theta}_{\hat{\mathcal{S}}}\right)\triangleq\left\Vert \boldsymbol{y}-\mathbf{A}_{\hat{\mathcal{S}}}\left(\boldsymbol{\theta}_{\hat{\mathcal{S}}}\right)\boldsymbol{x}_{\hat{\mathcal{S}}}\right\Vert ^{2}+C,\label{eq:theta_fun}
\end{equation}
where $C$ is a constant. However, it is difficult to find the optimal
$\boldsymbol{\bar{\theta}}_{\hat{\mathcal{S}}}$ that minimizes $\mathcal{L}\left(\boldsymbol{\theta}_{\hat{\mathcal{S}}}\right)$
since $\mathcal{L}\left(\boldsymbol{\theta}_{\hat{\mathcal{S}}}\right)$
is non-convex w.r.t. $\boldsymbol{\theta}_{\hat{\mathcal{S}}}$. To
address this challenge, we employ the BFGS algorithm, which is a quasi-Newton
method. The details of the BFGS algorithm are presented in the following
subsection.

\subsubsection{BFGS for grid refinement}

Unlike traditional gradient descent, the BFGS algorithm leverages
second-order derivative information of the objective function $\mathcal{L}\left(\boldsymbol{\theta}_{\hat{\mathcal{S}}}\right)$
with respect to $\boldsymbol{\theta}_{\hat{\mathcal{S}}}$ to determine
a more effective search direction \cite{numerical_opt}, which is
particularly advantageous for non-convex optimization problems.

The update rule of standard Newton's method for the grid parameters
$\boldsymbol{\theta}_{\hat{\mathcal{S}}}$ at the $i+1\textrm{-th}$
iteration is given by \cite{numerical_opt}:
\begin{align}
\boldsymbol{\theta}_{\hat{\mathcal{S}}}^{\left(i+1\right)} & =\boldsymbol{\theta}_{\hat{\mathcal{S}}}^{\left(i\right)}-\mathbf{F}_{i}^{-1}\nabla\mathcal{L}\left(\boldsymbol{\theta}_{\hat{\mathcal{S}}}^{\left(i\right)}\right),\label{eq:Newton_update}
\end{align}
where $\nabla\mathcal{L}\left(\boldsymbol{\theta}_{\hat{\mathcal{S}}}^{\left(i\right)}\right)$
is the gradient of $\mathcal{L}\left(\boldsymbol{\theta}_{\hat{\mathcal{S}}}\right)$
with respect to $\boldsymbol{\theta}_{\hat{\mathcal{S}}}$ at the
point $\boldsymbol{\theta}_{\hat{\mathcal{S}}}^{\left(i\right)}$,
and $\mathbf{F}_{i}$ is the Hessian matrix at the $i$-th iteration. 

However, computing the inverse of Hessian matrix at each iteration
is computationally prohibitive. For computational efficiency, the
BFGS algorithm avoids the direct computation of $\mathbf{F}_{i}^{-1}$
at each step but only maintains an approximation, denoted as $\mathbf{B}_{i}\approx\mathbf{F}_{i}^{-1}$.
Specially, the approximation of $\mathbf{B}_{i+1}$ is recursively
updated based on $\mathbf{B}_{i}$, using the following equation \cite{numerical_opt}:
\begin{equation}
\mathbf{B}_{i+1}=\left(\mathbf{I}_{\hat{\mathcal{S}}}-\rho_{i}\boldsymbol{p}_{i}\boldsymbol{q}_{i}^{T}\right)^{T}\mathbf{B}_{i}\left(\mathbf{I}_{\hat{\mathcal{S}}}-\rho_{i}\boldsymbol{p}_{i}\boldsymbol{q}_{i}^{T}\right)+\rho_{i}\boldsymbol{q}_{i}\boldsymbol{q}_{i}^{T}
\end{equation}
where $\boldsymbol{p}_{i}=\nabla\mathcal{L}\left(\boldsymbol{\theta}_{\hat{\mathcal{S}}}^{\left(i\right)}\right)-\nabla\mathcal{L}\left(\boldsymbol{\theta}_{\hat{\mathcal{S}}}^{\left(i-1\right)}\right)$
is the difference of gradient, $\boldsymbol{q}_{i}=\boldsymbol{\theta}_{\hat{\mathcal{S}}}^{\left(i\right)}-\boldsymbol{\theta}_{\hat{\mathcal{S}}}^{\left(i-1\right)}$
is the difference of grid parameter, and $\rho_{i}=\frac{1}{\boldsymbol{q}_{i}^{T}\boldsymbol{p}_{i}}$
is the scalar normalization factor.

For enhanced robustness, the descent direction is chosen adaptively
at each iteration. If the secant condition $\rho_{i}\geq0$ holds
\cite{numerical_opt}, it indicates the updated Hessian approximation
remains positive definite. In this case, the algorithm employs the
efficient quasi-Newton direction. Otherwise, the algorithm reverts
to the more conservative descent direction solely based on gradient.
As a result, the final update equation of BFGS is given by:
\begin{align}
\boldsymbol{\theta}_{\hat{\mathcal{S}}}^{\left(i+1\right)} & =\boldsymbol{\theta}_{\hat{\mathcal{S}}}^{\left(i\right)}+\epsilon_{\theta}^{\left(i\right)}\boldsymbol{d}^{\left(i\right)},\label{eq:final_BFGS_update}
\end{align}
where $\epsilon_{\theta}^{\left(i\right)}$ is the scalar step size
determined by Armijo rule, as detailed below. And the descent direction
$\boldsymbol{d}^{\left(i\right)}$ is chosen as:
\begin{equation}
\boldsymbol{d}^{\left(i\right)}=\begin{cases}
-\mathbf{B}_{i}\nabla\mathcal{L}\left(\boldsymbol{\theta}_{\hat{\mathcal{S}}}^{\left(i\right)}\right), & \textrm{if}\ \rho_{i}\geq0,\\
-\nabla\mathcal{L}\left(\boldsymbol{\theta}_{\hat{\mathcal{S}}}^{\left(i\right)}\right), & \textrm{otherwise}.
\end{cases}\label{eq:final_BFGS_descent}
\end{equation}

\subsubsection{Armijo rule for step-size calculation}

Once the BFGS algorithm provides the descent direction $\boldsymbol{d}^{\left(i\right)}$,
a backtrack line search that satisfies the Armijo rule is employed
to determine an appropriate step size $\epsilon_{\theta}^{\left(i\right)}$.
To simplify notation, the iteration index $i$ is omitted in the remaining
part of this subsection. Furthermore, to explicitly show the dependencies
of the objective functions, we denote $\mathcal{\psi}\left(\boldsymbol{x}_{\hat{\mathcal{S}}}\right)$
from (\ref{eq:x_fun}) as $\mathcal{\psi}\left(\boldsymbol{x}_{\hat{\mathcal{S}}};\boldsymbol{\theta}_{\hat{\mathcal{S}}}\right)$
and $\mathcal{L}\left(\boldsymbol{\theta}_{\hat{\mathcal{S}}}\right)$
from (\ref{eq:theta_fun}) as $\mathcal{L}\left(\boldsymbol{\theta}_{\hat{\mathcal{S}}};\boldsymbol{x}_{\hat{\mathcal{S}}}\right)$.

For a given $\boldsymbol{\theta}_{\hat{\mathcal{S}}}$ and a chosen
hyper-parameter $c\in\left[0,1\right]$, the step size $\epsilon_{\theta}$
satisfies the following inequality \cite{numerical_opt}:
\begin{align}
\mathcal{L}\left(\boldsymbol{\theta}_{\hat{\mathcal{S}}}+\epsilon_{\theta}\boldsymbol{d};\boldsymbol{x}_{\hat{\mathcal{S}}}^{\text{opt}}\left(\boldsymbol{\theta}_{\hat{\mathcal{S}}}+\epsilon_{\theta}\boldsymbol{d}\right)\right)\leq & \mathcal{L}\left(\boldsymbol{\theta}_{\hat{\mathcal{S}}};\boldsymbol{x}_{\hat{\mathcal{S}}}^{\text{opt}}\left(\boldsymbol{\theta}_{\hat{\mathcal{S}}}\right)\right)\nonumber \\
+ & c\epsilon_{\theta}\boldsymbol{d}^{T}\nabla\mathcal{L}\left(\boldsymbol{\theta}_{\hat{\mathcal{S}}};\boldsymbol{x}_{\hat{\mathcal{S}}}^{\text{opt}}\left(\boldsymbol{\theta}_{\hat{\mathcal{S}}}\right)\right)\label{eq:armijo_rule}
\end{align}
where the optimal gain vector is defined as $\boldsymbol{x}_{\hat{\mathcal{S}}}^{\text{opt}}\left(\boldsymbol{\theta}_{\hat{\mathcal{S}}}\right)=\arg\min_{\boldsymbol{x}_{\hat{\mathcal{S}}}}\ \mathcal{\psi}\left(\boldsymbol{x}_{\hat{\mathcal{S}}};\boldsymbol{\theta}_{\hat{\mathcal{S}}}\right)$.
The hyper-parameter $c$ enforces the sufficient decrease condition,
and is typically set to a small positive constant such as $10^{-2}$.
The line search is implemented by starting with an initial step size
$\epsilon_{0}$, and iteratively reducing it by a backtracking factor
$\gamma\in\left[0,1\right]$, until the condition is met.

It is noteworthy that our step-size calculation scheme leverages the
principle of the variable projection \cite{variable_projection}.
Specifically, our method requires recomputing the corresponding optimal
channel gain vector $\boldsymbol{x}_{\hat{\mathcal{S}}}^{\text{opt}}$,
since $\boldsymbol{x}_{\hat{\mathcal{S}}}^{\text{opt}}$ is implicitly
determined by $\boldsymbol{\theta}_{\hat{\mathcal{S}}}$ at every
point. In other words, the calculation of the step size is purely
decided by $\boldsymbol{\theta}_{\hat{\mathcal{S}}}$ and does not
depend on a fixed channel gain vector, which is different from a simple
alternating optimization scheme. While the variable projection approach
incurs a higher computational cost during the process of line search,
it facilitates a more robust and effective search, leading to faster
overall convergence and performance, as validated in Section \ref{sec:Simulations}.

\section{Sparse Signal Estimation Module in Slow Timescale\label{sec:Turbo-VBI-Algorithm}}

\subsection{Tanh-VBI Algorithm Based on the Mean Field VBI Framework}

Given the observation $\boldsymbol{y}$, the SSE module adopts the
mean field VBI \cite{Tzikas_VBI} to calculate the approximate marginal
posteriors $q\left(\boldsymbol{v}\right)$ for fixed grid parameters
$\hat{\boldsymbol{\theta}}$. Since the grid parameter is fixed in
the SSE module, we omit the grid parameter $\hat{\boldsymbol{\theta}}$
and use $\mathbf{A}$ as a simplified notation for $\mathbf{A}\left(\hat{\boldsymbol{\theta}}\right)$. 

We first give an overview of the mean field variational Bayesian inference
before presenting the update equations in Tanh-VBI. For convenience,
we use $\boldsymbol{v}^{k}$ to denote an individual variable in $\boldsymbol{v}\triangleq\left\{ \boldsymbol{x},\boldsymbol{\rho},\boldsymbol{s},\kappa\right\} $
and let $\mathcal{H}=\{k\mid\forall\boldsymbol{v}^{k}\in\boldsymbol{v}\}$.We
aim at calculating the posterior distribution of random variables
with the prior $p\left(\boldsymbol{x},\boldsymbol{\rho},\boldsymbol{s}\right)$
in (\ref{eq:p(x,rou,s)}), i.e., $p\left(\boldsymbol{v}\mid\boldsymbol{y}\right)$.
However, it is usually intractable to find the posterior directly
since the considered problem involves integrals of many high-dimensional
variables. Based on the mean field VBI method, the approximate marginal
posterior could be calculated by minimizing the KLD between $p\left(\boldsymbol{v}\mid\boldsymbol{y}\right)$
and $q\left(\boldsymbol{v}\right)$, subject to a factorized form
constraint as \cite{Tzikas_VBI}:
\begin{eqnarray}
\mathcal{\mathscr{A}}_{\mathrm{VBI}}: & q^{*}\left(\boldsymbol{v}\right) & =\arg\min_{q\left(\boldsymbol{v}\right)}\int q\left(\boldsymbol{v}\right)\ln\frac{q\left(\boldsymbol{v}\right)}{p\left(\boldsymbol{v}\mid\boldsymbol{y}\right)}\textrm{d}\boldsymbol{v},\label{eq:KLDmin}\\
\mathrm{s.t.} & q\left(\boldsymbol{v}\right) & =\prod_{k\in\mathcal{H}}q\left(\boldsymbol{v}^{k}\right),\int q\left(\boldsymbol{v}^{k}\right)d\boldsymbol{v}^{k}=1,\label{eq:factorconstrain}
\end{eqnarray}
where the constraint $q\left(\boldsymbol{v}\right)=\prod_{k\in\mathcal{H}}q\left(\boldsymbol{v}^{k}\right)$
is the mean field assumption \cite{Tzikas_VBI}.

Although the problem $\mathcal{\mathscr{A}}_{\mathrm{VBI}}$ is known
to be non-convex, it is convex w.r.t. a single variational distribution
$q\left(\boldsymbol{v}^{l}\right)$ after fixing other variational
distributions $q\left(\boldsymbol{v}^{k}\right),\forall k\neq l$
\cite{Tzikas_VBI}. And it has been proved in \cite{Tzikas_VBI} that
a stationary solution could be found via optimizing each variational
distribution in an alternating fashion. Specifically, for given $q\left(\boldsymbol{v}^{k}\right),\forall k\neq l$,
the optimal $q\left(\boldsymbol{v}^{l}\right)$ that minimizes the
KL-divergence is given by \cite{Tzikas_VBI}:
\begin{equation}
q\left(\boldsymbol{v}^{l}\right)=\frac{\exp\left(\left\langle \ln p\left(\boldsymbol{v},\boldsymbol{y}\right)\right\rangle _{\Pi_{k\neq l}q\left(\boldsymbol{v}^{k}\right)}\right)}{\int\exp\left(\left\langle \ln p\left(\boldsymbol{v},\boldsymbol{y}\right)\right\rangle _{\Pi_{k\neq l}q\left(\boldsymbol{v}^{k}\right)}\right)\textrm{d}\boldsymbol{v}^{l}},\label{eq:q(vl)}
\end{equation}
where $\left\langle \cdot\right\rangle _{\Pi_{k\neq l}q\left(\boldsymbol{v}^{k}\right)}$
is an expectation operation w.r.t. $q\left(\boldsymbol{v}^{k}\right)$
for $k\neq l$. The joint distribution $p\left(\boldsymbol{v},\boldsymbol{y}\right)$
is given by
\begin{align}
p\left(\boldsymbol{v},\boldsymbol{y}\right) & =p\left(\boldsymbol{y}\mid\boldsymbol{x},\kappa\right)p\left(\boldsymbol{x},\boldsymbol{\rho},\boldsymbol{s}\right)p\left(\kappa\right),\label{eq:joint distribution}
\end{align}
where $p\left(\boldsymbol{y}\mid\boldsymbol{x},\kappa\right)=\mathcal{CN}\left(\boldsymbol{y};\mathbf{A}\boldsymbol{x},\kappa^{-1}\mathbf{I}_{M}\right)$
is the likelihood function, $p\left(\boldsymbol{x},\boldsymbol{\rho},\boldsymbol{s}\right)$
and $p\left(\kappa\right)$ are the priors given in (\ref{eq:p(x,rou,s)})
and (\ref{eq:p(gamma)}), respectively. By finding a stationary solution
$q^{*}\left(\boldsymbol{v}\right)$ of $\mathcal{\mathscr{A}}_{\mathrm{VBI}}$,
we could obtain the approximate posterior $q^{*}\left(\boldsymbol{v}^{k}\right)\thickapprox p\left(\boldsymbol{v}^{k}\mid\boldsymbol{y}\right)$.

By substituting the joint distribution (\ref{eq:joint distribution})
into (\ref{eq:q(vl)}), each optimal variational distribution $q\left(\boldsymbol{v}^{l}\right)$
can be derived. In the following, we start with providing a detailed
derivation for $q\left(\boldsymbol{x}\right)$, since the primary
analytical challenge of our proposed algorithm arises from this update
step. 

\subsection{Update Equation of $q\left(\boldsymbol{x}\right)$}

For given $q\left(\boldsymbol{\rho}\right)$, $q\left(\boldsymbol{s}\right)$
and $q\left(\kappa\right)$, the posterior distribution $q\left(\boldsymbol{x}\right)$
can be expressed as:
\begin{eqnarray}
\textrm{ln}q\left(\boldsymbol{x}\right) & \propto & \left\langle \ln\left(p\left(\boldsymbol{v},\boldsymbol{y}\right)\right)\right\rangle _{q\left(\boldsymbol{\rho}\right)q\left(\boldsymbol{s}\right)q\left(\kappa\right)}\nonumber \\
 & \propto & \left\langle \ln\left(p\left(\boldsymbol{y}\mid\boldsymbol{x},\kappa\right)\right)\right\rangle _{q\left(\kappa\right)}+\left\langle \ln\left(p\left(\boldsymbol{x},\boldsymbol{\rho},\boldsymbol{s}\right)\right)\right\rangle _{q\left(\boldsymbol{\rho}\right)q\left(\boldsymbol{s}\right)}\nonumber \\
 & \propto & \left\langle \ln\left(p\left(\boldsymbol{y}\mid\boldsymbol{x},\kappa\right)\right)\right\rangle _{q\left(\kappa\right)}+\left\langle \ln\left(p\left(\boldsymbol{x}\mid\boldsymbol{\rho}\right)\right)\right\rangle _{q\left(\boldsymbol{\rho}\right)}\nonumber \\
 & \propto & \underbrace{-\left\langle \kappa\right\rangle \mid\mid\boldsymbol{y}-\mathbf{A}\boldsymbol{x}\mid\mid^{2}}_{\textrm{likelihood}}+\underbrace{\left\langle \ln\left(p\left(\boldsymbol{x}\mid\boldsymbol{\rho}\right)\right)\right\rangle _{q\left(\boldsymbol{\rho}\right)}}_{\textrm{prior}},\label{eq:ln_q(x)}
\end{eqnarray}
where $p\left(\boldsymbol{x}\mid\boldsymbol{\rho}\right)$ is modeled
as a product of element-wise tanh distributions in (\ref{eq:element_wise_tanh}).
Consequently, $\left\langle \ln\left(p\left(\boldsymbol{x}\mid\boldsymbol{\rho}\right)\right)\right\rangle _{q\left(\boldsymbol{\rho}\right)}$
can be expressed as:
\begin{eqnarray}
\left\langle \ln\left(p\left(\boldsymbol{x}\mid\boldsymbol{\rho}\right)\right)\right\rangle _{q\left(\boldsymbol{\rho}\right)} & \propto & \int\ln\left(p\left(\boldsymbol{x}\mid\boldsymbol{\rho}\right)\right)q\left(\boldsymbol{\rho}\right)d\boldsymbol{\rho},\nonumber \\
 & \propto & -\sum_{n}\int q\left(\rho_{n}\right)\rho_{n}\textrm{tanh}\left(\frac{\mid x_{n}\mid^{2}}{\zeta}\right)d\rho_{n}\nonumber \\
 & = & -\sum_{n}\left\langle \rho_{n}\right\rangle \textrm{tanh}\left(\frac{\mid x_{n}\mid^{2}}{\zeta}\right),\label{eq:multi_tanh}
\end{eqnarray}

Substituting \eqref{eq:multi_tanh} into (\ref{eq:ln_q(x)}), $\textrm{ln}q\left(\boldsymbol{x}\right)$
can be expressed as:
\begin{align}
\textrm{ln}q\left(\boldsymbol{x}\right)\propto & -\left\langle \kappa\right\rangle \mid\mid\boldsymbol{y}-\mathbf{A}\boldsymbol{x}\mid\mid^{2}\nonumber \\
- & \sum_{n}\left\langle \rho_{n}\right\rangle \textrm{tanh}\left(\frac{\mid x_{n}\mid^{2}}{\zeta}\right).\label{eq:refined_ln_q(x)}
\end{align}

It is noteworthy that the second term in (\ref{eq:refined_ln_q(x)})
involves the nonlinear function $\textrm{tanh}\left(\cdotp\right)$,
which makes it difficult to directly obtain the closed-form of $\textrm{ln}q\left(\boldsymbol{x}\right)$.
Consequently, the expectations required to update the other variables
cannot be computed directly. To \textcolor{black}{solve this challenge,
we approximate} $q\left(\boldsymbol{x}\right)$ \textcolor{black}{as
a Gaussian distribution based on a successive linear approximation
(SLA) method, as will be detailed in the next subsection.}

\subsection{Successive Linear Approximation \label{sec:SLA}}

The core idea of SLA is to linearize the non-linear function around
a specific point. For each iteration, we define $\hat{\boldsymbol{u}}$
as the posterior mean of $\boldsymbol{x}$ obtained from the previous
iteration, and $\hat{u}_{n}$ is the $n$-th element of $\hat{\boldsymbol{u}}$,
such that nonlinear $\textrm{tanh}\left(z\right)$ with $z=\frac{\mid x_{n}\mid^{2}}{\zeta}$
can be approximated using a first-order Taylor expansion around the
point $z_{0}=\frac{\mid\hat{u}_{n}\mid^{2}}{\zeta}$ :
\begin{align}
\textrm{tanh}\left(\frac{\mid x_{n}\mid^{2}}{\zeta}\right)\thickapprox & \textrm{tanh}\left(\frac{\mid\hat{u}_{n}\mid^{2}}{\zeta}\right)\nonumber \\
+ & \frac{\partial\tanh\left(z\right)}{\partial z}\bigg|_{z=\frac{\mid\hat{u}_{n}\mid^{2}}{\zeta}}\left(\frac{\mid x_{n}\mid^{2}}{\zeta}-\frac{\mid\hat{u}_{n}\mid^{2}}{\zeta}\right)\nonumber \\
= & \hat{a}_{n}+\hat{b}_{n}\frac{\mid x_{n}\mid^{2}}{\zeta},\label{eq:linear_tanh}
\end{align}
where $\frac{\partial\tanh\left(z\right)}{\partial z}=1-\tanh^{2}\left(z\right)$,
and $\hat{b}_{n}$ and $\hat{a}_{n}$ are the slope and intercept
of this linear approximation, respectively. They are treated as constants
within the current iteration, and can be expressed as:
\begin{align}
\hat{b}_{n} & =1-\tanh^{2}\left(\frac{\mid\hat{u}_{n}\mid^{2}}{\zeta}\right),\nonumber \\
\hat{a}_{n}= & \textrm{tanh}\left(\frac{\mid\hat{u}_{n}\mid^{2}}{\zeta}\right)-\hat{b}_{n}\frac{\mid\hat{u}_{n}\mid^{2}}{\zeta}.
\end{align}

By substituting \eqref{eq:linear_tanh} into (\ref{eq:refined_ln_q(x)}),
the log-posterior $\textrm{ln}q\left(\boldsymbol{x}\right)$ can be
simplified as:
\begin{eqnarray}
\textrm{ln}q\left(\boldsymbol{x}\right) & \thickapprox & -\left\langle \kappa\right\rangle \mid\mid\boldsymbol{y}-\mathbf{A}\boldsymbol{x}\mid\mid^{2}-\sum_{n}\left\langle \rho_{n}\right\rangle \left(\hat{a}_{n}+\hat{b}_{n}\frac{\mid x_{n}\mid^{2}}{\zeta}\right)\nonumber \\
 & \propto & -\left\langle \kappa\right\rangle \mid\mid\boldsymbol{y}-\mathbf{A}\boldsymbol{x}\mid\mid^{2}-\sum_{n}\left\langle \rho_{n}\right\rangle \frac{\hat{b}_{n}}{\zeta}\mid x_{n}\mid^{2}\nonumber \\
 & = & -\left\langle \kappa\right\rangle \mid\mid\boldsymbol{y}-\mathbf{A}\boldsymbol{x}\mid\mid^{2}-\boldsymbol{x}^{H}\textrm{diag}\left(\boldsymbol{c}\right)\boldsymbol{x}
\end{eqnarray}
where $\boldsymbol{c}$ can be expressed as:
\begin{equation}
\boldsymbol{c}=\frac{\left\langle \boldsymbol{\rho}\right\rangle }{\zeta}\left(1-\tanh^{2}\left(\frac{\mid\hat{\boldsymbol{u}}\mid^{2}}{\zeta}\right)\right)\label{eq:c_n}
\end{equation}

After applying this approximation, the log-posterior $\textrm{ln}q\left(\boldsymbol{x}\right)$
exhibits a quadratic form in $\boldsymbol{x}$, which implies that
$q\left(\boldsymbol{x}\right)$ can be expressed as a Gaussian distribution,
similar to the original VBI based on a Bernoulli-Gamma-Gaussian (BGG)
prior model in \cite{Liu_turbo_VBI}:
\begin{equation}
q\left(\boldsymbol{x}\right)=\mathcal{CN}\left(\boldsymbol{x};\boldsymbol{\mu}_{\textrm{tanh}},\mathbf{\Sigma}_{\textrm{tanh}}\right),\label{eq:SLA_x}
\end{equation}
where the approximate posterior parameters are given by:
\begin{align}
\mathbf{\Sigma}_{\textrm{tanh}}= & \left(\textrm{diag}\left(\boldsymbol{c}\right)+\left\langle \kappa\right\rangle \mathbf{A}^{H}\mathbf{A}\right)^{-1},\label{eq:SLA_x_para}\\
\boldsymbol{\mu}_{\textrm{tanh}}= & \mathbf{\Sigma}\mathbf{A}^{H}\left\langle \kappa\right\rangle \boldsymbol{y}.\nonumber 
\end{align}

It is worth noting that while our algorithm approximates $q\left(\boldsymbol{x}\right)$
with a Gaussian distribution, it differs fundamentally from the original
VBI algorithm. The key distinction lies in how the posterior covariance
is determined. In the original VBI, the posterior covariance matrix
is static for a given hyper-parameter $\left\langle \boldsymbol{\rho}\right\rangle $.
However, in our proposed method, the posterior covariance matrix is
adaptive. As shown in (\ref{eq:SLA_x_para}), it depends on the term
$\textrm{diag}\left(\boldsymbol{c}\right)$, which is updated in each
iteration based on the posterior mean $\hat{\boldsymbol{u}}$ obtained
from the previous iteration. This adaptive process allows our proposed
algorithm to model the true non-Gaussian posterior more precisely
than methods that rely on static parameters.

\subsection{Update Equation of Other Variables \label{sec:other_variable_equation}}

With the variational posterior distribution $q\left(\boldsymbol{x}\right)$
approximated as a Gaussian distribution $\mathcal{CN}\left(\boldsymbol{x};\boldsymbol{\mu}_{\textrm{tanh}},\mathbf{\Sigma}_{\textrm{tanh}}\right)$
via the SLA method, the update equations for the remaining variational
distributions, i.e, $q\left(\boldsymbol{\rho}\right)$, $q\left(\boldsymbol{s}\right)$,
$q\left(\kappa\right)$, are similar to that in \cite{Liu_turbo_VBI}.
For the sake of brevity, we directly present the final update equations
in this section.

\subsubsection{Update of $q\left(\boldsymbol{\rho}\right)$}

The posterior distribution $q\left(\boldsymbol{\rho}\right)$ can
be computed by
\begin{equation}
q\left(\boldsymbol{\rho}\right)=\prod_{n=1}^{N}\textrm{Ga}\left(\rho_{n};\widetilde{a}_{n},\widetilde{b}_{n}\right),\label{eq:q(rou)}
\end{equation}
where the parameters $\widetilde{a}_{n}$ and $\widetilde{b}_{n}$
are given by 
\begin{equation}
\begin{aligned}\widetilde{a}_{n} & =\left\langle s_{n}\right\rangle a_{n}+\left\langle 1-s_{n}\right\rangle \overline{a}_{n}+1,\\
\widetilde{b}_{n} & =\left\langle s_{n}\right\rangle b_{n}+\left\langle 1-s_{n}\right\rangle \overline{b}_{n}+\left|\boldsymbol{\mu}_{\textrm{tanh}}^{n}\right|^{2}+\mathbf{\Sigma}_{\textrm{tanh}}^{n}.
\end{aligned}
\label{eq:rou_para}
\end{equation}
where $\boldsymbol{\mu}_{\textrm{tanh}}^{n}$ is the $n\textrm{-th}$
element of $\boldsymbol{\mu}_{\textrm{tanh}}$, and $\mathbf{\Sigma}_{\textrm{tanh}}^{n}$
is the $n\textrm{-th}$ diagonal element of $\mathbf{\Sigma}_{\textrm{tanh}}$.

\subsubsection{Update of $q\left(\boldsymbol{s}\right)$}

The posterior distribution $q\left(\boldsymbol{s}\right)$ can be
calculated by
\begin{equation}
q\left(\boldsymbol{s}\right)=\prod_{n=1}^{N}\left(\widetilde{\lambda}_{n}\right)^{s_{n}}\left(1-\widetilde{\lambda}_{n}\right)^{1-s_{n}},\label{eq:q(s)}
\end{equation}
where $\widetilde{\lambda}_{n}$ is given by
\begin{equation}
\widetilde{\lambda}_{n}=\frac{\lambda_{n}C_{n}}{\lambda_{n}C_{n}+\left(1-\lambda_{n}\right)\overline{C}_{n}},\label{eq:s_para}
\end{equation}
with $C_{n}=\dfrac{b_{n}^{a_{n}}}{\Gamma\left(a_{n}\right)}\exp\left(\left(a_{n}-1\right)\left\langle \ln\rho_{n}\right\rangle -b_{n}\left\langle \rho_{n}\right\rangle \right)$
and $\overline{C}_{n}=\dfrac{\overline{b}_{n}^{\overline{a}_{n}}}{\Gamma\left(\overline{a}_{n}\right)}\exp\left(\left(\overline{a}_{n}-1\right)\left\langle \ln\rho_{n}\right\rangle -\overline{b}_{n}\left\langle \rho_{n}\right\rangle \right)$.
Here, $\Gamma\left(\cdot\right)$ denotes the gamma function.

\subsubsection{Update of $q\left(\kappa\right)$}

The posterior distribution $q\left(\kappa\right)$ is given by
\begin{equation}
q\left(\kappa\right)=\textrm{Ga}\left(\kappa;\widetilde{c},\widetilde{d}\right),\label{eq:q(gamma)}
\end{equation}
where the parameters $\widetilde{c}$ and $\widetilde{d}$ are given
by
\begin{equation}
\begin{aligned}\widetilde{c} & =c+M,\\
\widetilde{d} & =d+\left\Vert \boldsymbol{y}-\mathbf{A}\left(\boldsymbol{\theta}\right)\boldsymbol{\mu}_{\textrm{tanh}}\right\Vert ^{2}+\textrm{tr}\left(\mathbf{A}\left(\boldsymbol{\theta}\right)\mathbf{\Sigma}_{\textrm{tanh}}\mathbf{A}\left(\boldsymbol{\theta}\right)^{H}\right).
\end{aligned}
\label{eq:gamma_para}
\end{equation}

It is important to note that the calculation of $\widetilde{d}$ contains
the term $\textrm{tr}\left(\mathbf{A}\left(\boldsymbol{\theta}\right)\mathbf{\Sigma}_{\textrm{tanh}}\mathbf{A}\left(\boldsymbol{\theta}\right)^{H}\right)$,
which involves large-scale matrix multiplications. For high-dimensional
problems, this step can become a significant computational bottleneck.
To enhance computational efficiency, a common and effective approximation
is to only consider the diagonal elements of the covariance matrix
$\mathbf{\Sigma}_{\textrm{tanh}}$. Under this diagonal approximation,
the trace term can be simplified to a much more efficient computation:
\begin{equation}
\textrm{tr}\left(\mathbf{A}\left(\boldsymbol{\theta}\right)\mathbf{\Sigma}_{\textrm{tanh}}\mathbf{A}\left(\boldsymbol{\theta}\right)^{H}\right)=\sum_{n}\sigma_{\textrm{tanh}}^{2}\left\Vert \boldsymbol{a}_{n}\left(\boldsymbol{\theta}\right)\right\Vert ^{2},\label{eq:var_simplify}
\end{equation}
where $\sigma_{\textrm{tanh}}^{2}$ is the $n$-th diagonal element
of $\mathbf{\Sigma}_{\textrm{tanh}}$, and $\boldsymbol{a}_{n}\left(\boldsymbol{\theta}\right)$
is the $n$-th column of the matrix $\mathbf{A}\left(\boldsymbol{\theta}\right)$.
This reduces the complexity to a simple vector norm calculation, dramatically
reducing the complexity.

Finally, the expectations used in the above update expressions are
summarized as follows:
\[
\begin{aligned}\left\langle \rho_{n}\right\rangle = & \begin{aligned}\dfrac{\widetilde{a}_{n}}{\widetilde{b}_{n}},\left\langle \boldsymbol{\rho}\right\rangle =\left[\bigl\langle\rho_{1}\bigr\rangle,\ldots,\bigl\langle\rho_{N}\bigr\rangle\right]^{T},\left\langle s_{n}\right\rangle =\widetilde{\lambda}_{n},\end{aligned}
\\
\left\langle \kappa\right\rangle =\dfrac{\widetilde{c}}{\widetilde{d}} & \begin{aligned},\left\langle \ln\rho_{n}\right\rangle =\psi\left(\widetilde{a}_{n}\right)-\ln\widetilde{b}_{n},\end{aligned}
\end{aligned}
\]
where $\psi\left(\cdot\right)\triangleq d\ln\left(\Gamma\left(\cdot\right)\right)$
denotes the logarithmic derivative of the gamma function.

\subsection{Algorithm Summary and Complexity Analysis \label{sec:complexity}}

The overall two-timescale alternating MAP framework, with the tanh-VBI
algorithm and BFGS method as its core, is summarized in Algorithm
\ref{AE-SC-VBI}.

\begin{algorithm}[t]
\begin{singlespace}
{\small\caption{\label{AE-SC-VBI}The proposed two-timescale alternating MAP framework.}
}{\small\par}

\textbf{Input:} Received signal $\boldsymbol{y}$, initial dense grid
$\boldsymbol{\theta}$ and corresponding sensing matrix $\mathbf{A}\left(\boldsymbol{\theta}\right)$,
maximum iteration numbers $I_{0}$, $I_{1}$, $I_{2}$.

\textbf{Output:} Estimated sparse signal $\hat{\boldsymbol{x}}$,
estimated support $\hat{\mathcal{S}}$, and active grid $\hat{\boldsymbol{\theta}}_{\hat{\mathcal{S}}}$.

\begin{algorithmic}[1]

\FOR{${\color{blue}{\color{black}t=1,\cdots,I_{0}}}$}

\STATE \textbf{Sparse Signal Estimation (SSE) Module at Slow Timescale:}

\STATE Initialize the distribution functions $q\left(\boldsymbol{s}\right)$,
$q\left(\boldsymbol{\rho}\right)$ and $q\left(\kappa\right)$.

\FOR{${\color{blue}{\color{black}k=1,\cdots,I_{1}}}$}

\STATE Update $q^{k}\left(\boldsymbol{x}\right)$ using (\ref{eq:SLA_x})-(\ref{eq:SLA_x_para}).
The vector $\boldsymbol{c}$ within the posterior covariance matrix
$\mathbf{\Sigma}_{\textrm{tanh}}$ is obtained by (\ref{eq:c_n}),
which depends on the posterior mean $\hat{\boldsymbol{u}}$ from the
previous iteration.

\STATE Update $q^{k}\left(\boldsymbol{\rho}\right)$ using (\ref{eq:q(rou)})-(\ref{eq:rou_para}).

\STATE Update $q^{k}\left(\boldsymbol{s}\right)$ using (\ref{eq:q(s)})-(\ref{eq:s_para}).

\STATE Update $q^{k}\left(\kappa\right)$ using (\ref{eq:q(gamma)})-(\ref{eq:var_simplify}).

\ENDFOR

\STATE Obtain the MAP estimators $\hat{\boldsymbol{x}}$ and $\hat{\kappa}$
of $\boldsymbol{x}$ and $\kappa$ from the estimated $\hat{p}\left(\boldsymbol{x}\mid\boldsymbol{y}\right)=q^{I_{1}}\left(\boldsymbol{x}\right)$
and $\hat{p}\left(\kappa\mid\boldsymbol{y}\right)=q^{I_{1}}\left(\kappa\right)$,
and calculate the estimated support $\hat{\mathcal{S}}$ from $\hat{\boldsymbol{x}}$.

\STATE\textbf{ Super-Resolution Grid Update (SR-GU) Module at Fast
Timescale:}

\FOR{${\color{blue}{\color{black}j=1,\cdots,I_{2}}}$}

\STATE Given fixed $\boldsymbol{\theta}_{\hat{\mathcal{S}}}$, construct
the MAP optimization problem of $\boldsymbol{x}_{\hat{\mathcal{S}}}$
using (\ref{eq:x_fun}).

\STATE Obtain the MAP estimator $\boldsymbol{x}_{\hat{\mathcal{S}}}$
by performing the LMMSE method in (\ref{LMMSE}).

\STATE Given fixed $\boldsymbol{x}_{\hat{\mathcal{S}}}$, construct
the ML optimization problem of $\boldsymbol{\theta}_{\hat{\mathcal{S}}}$
using (\ref{eq:theta_fun}).

\STATE Obtain the ML estimator $\hat{\boldsymbol{\theta}}$ of $\boldsymbol{\theta}$
by performing the BFGS method using (\ref{eq:Newton_update})-(\ref{eq:final_BFGS_descent}),
and the step size is calculated by Armijo rule in (\ref{eq:armijo_rule}).

\ENDFOR

\ENDFOR

\STATE Output $\hat{\boldsymbol{x}}$, $\hat{\mathcal{S}}$ and $\hat{\boldsymbol{\theta}}_{\hat{\mathcal{S}}}$.

\end{algorithmic}
\end{singlespace}
\end{algorithm}

We further demonstrate the complexity of the proposed two-timescale
alternating MAP framework. For convenience, we define the number of
outer loop iterations as $I_{0}$, the number of inner iterations
for the tanh-VBI algorithm in the SSE module as $I_{1}$, and the
number of inner iterations for the SR-GU module as $I_{2}$. Recall
that the dimension of the sensing matrix $\mathbf{A}\left(\boldsymbol{\theta}\right)$
is $M\times N$.

The primary computational complexity comes from the SSE module, which
executes the tanh-VBI algorithm once in each outer-loop iteration.
Its complexity per outer iteration is $\mathcal{O}\left(I_{1}N^{3}\right)$,
which comprises $I_{1}$ internal updates, each requiring an $N$-dimensional
matrix inversion for the posterior update of $q\left(\boldsymbol{x}\right)$.
It is worth noting that the complexity can be further reduced from
$\mathcal{O}\left(I_{1}N^{3}\right)$ to $\mathcal{O}\left(I_{1}MN\right)$
by replacing the matrix inversion step with state-of-the-art inverse-free
algorithms, as shown in \cite{IF_VBI,SC_VBI}. 

In contrast, the SR-GU module executes a fast timescale refinement
on the active dynamic grid $\boldsymbol{\theta}_{\hat{\mathcal{S}}}$,
which has a dimension of $S=\left|\hat{\mathcal{S}}\right|\ll M$.
The complexity of this module is $\mathcal{O}\left(I_{2}MS^{2}\right)$,
resulting from $I_{2}$ alternating updates of $\boldsymbol{\theta}_{\hat{\mathcal{S}}}$
and its corresponding gain vector $\boldsymbol{x}_{\hat{\mathcal{S}}}$.

By combining the costs of both modules over the $I_{0}$ outer loop
iterations, the total computational complexity of the proposed framework
can be formulated. The complexity order is $\mathcal{O}\left(I_{0}\left(I_{1}N^{3}+I_{2}MS^{2}\right)\right)$,
where $I_{2}$ is chosen to be larger than $I_{1}$ in practice. This
highlights the two-timescale nature of the algorithm, which enables
more accurate grid refinement on a faster timescale.

It is noted that the complexity of baseline schemes, such as QNOMP
and DMRA, is not analytically derived here. This is because these
methods often lack a unified algorithmic framework suitable for direct
comparison. Instead, we provide a practical comparison of the computational
time between our proposed algorithm and various baselines in the simulation
section, as presented in \ref{sec:Simulations}.

\section{Simulations \label{sec:Simulations}}

In this section, we use the massive MIMO channel extrapolation problem
described in Section \ref{sec:System-Model} as an example to demonstrate
the advantages of the proposed two-timescale alternating MAP framework
approach. The baseline algorithms considered in the simulations are
described below.
\begin{itemize}
\item \textbf{ST-MUSIC aided Turbo-CS (MUSIC-CS) \cite{WAN_TWC}:} The ST-MUSIC
algorithm is first applied to obtain high-resolution estimation of
the grid parameters required for constructing the sensing matrix.
Subsequently, an EM-based Turbo-CS algorithm is used for channel extrapolation.
\item \textbf{QNOMP} \textbf{\cite{QNOMP}:} An initial on-grid OMP estimation
is performed to identify dominant components, which are then refined
through off-grid quasi-Newton BFGS optimization to further improve
accuracy.
\item \textbf{DMRA \cite{DMRA}:} A smooth relaxation based on the tanh
function is introduced to encourage sparsity, enabling joint estimation
of dominant grid components and their associated complex gains for
super-resolution channel extrapolation.
\end{itemize}

\subsection{Implementation Details}

In the simulations, the BS is equipped with a ULA consisting of $N_{r}=256$
antennas. Each BWP contains $M=100$ subcarriers, with each subcarrier
occupying a bandwidth of $f_{0}=120$ KHz, and the overall system
bandwidth is partitioned into $h_{p}=4$ BWPs. The uplink pilot vector
$\boldsymbol{\beta}$ in (\ref{eq:received signal model}) is generated
as a complex random vector, where each element has a random phase
and unit modulus. 

Note that in channel extrapolation problem, the delay and angle separations
among two nearest propagation paths is usually smaller than the resolution
determined by the DFT, which is typically employed for parameter estimation
\cite{DFT_baseline}. As such, the algorithm should have super resolution
capability in order to achieve a good channel extrapolation performance.

The normalized mean square error (NMSE) is used as the performance
metric for channel extrapolation, which is defined as:
\[
\textrm{NMSE\ensuremath{\left(\textrm{dB}\right)}}=10\log_{10}\frac{\left\Vert \widehat{\boldsymbol{h}}-\boldsymbol{h}\right\Vert ^{2}}{\left\Vert \boldsymbol{h}\right\Vert ^{2}}
\]
where $\widehat{\boldsymbol{h}}$ and $\boldsymbol{h}$ represents
the estimated and true fullband channels, respectively. Note that
the channel extrapolation performance directly reflects the estimation
accuracy of both angles and delays, as accurate parameter estimation
yields improved channel extrapolation performance.
\begin{figure}
\centering{}\includegraphics[width=75mm]{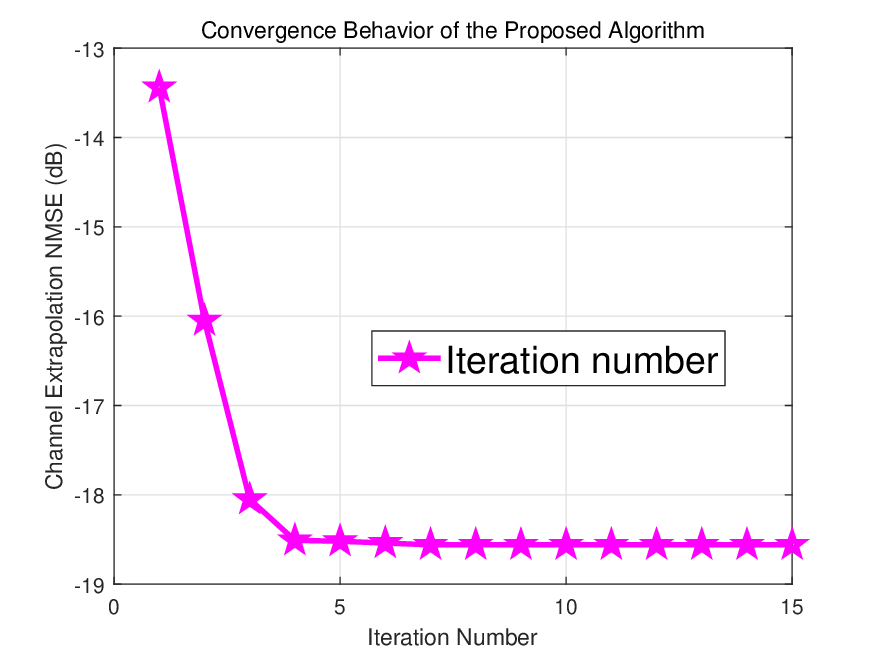}\caption{\label{fig:overall_convergence}{\small NMSE of channel extrapolation
versus the number of iterations}.}
\end{figure}

Furthermore, to directly quantify the estimation accuracy of angle
and delay parameters, we also evaluate the root mean square error
(RMSE) of the parameter estimation. To handle the different physical
units of angles and delays, their estimation errors are individually
normalized. Specifically, we introduce normalization factors $C_{\theta}$
and $C_{\tau}$ representing the dynamic range of the true angles
and delays, respectively. The normalized RMSE is then defined as:
\[
\textrm{RMSE\ensuremath{\left(\textrm{dB}\right)}}=10\log_{10}\frac{1}{K_{\textrm{est}}}\sum_{k}^{K_{\textrm{est}}}\left[\left(\frac{\Delta_{\theta,k}}{C_{\theta}}\right)^{2}+\left(\frac{\Delta_{\tau,k}}{C_{\tau}}\right)^{2}\right]
\]
where $K_{\textrm{est}}$ is the number of estimated paths, $\Delta_{\theta,k}=\theta_{k,\textrm{est}}-\theta_{k,\textrm{real}}$
is the error of angle estimation, $\Delta_{\tau,k}=\tau_{k,\textrm{est}}-\tau_{k,\textrm{real}}$
is the error of delay estimation, $\theta_{k,\textrm{est}}$, $\tau_{k,\textrm{est}}$
is the $k$-th estimated angle-delay pair, and $\theta_{k,\textrm{real}}$,
$\tau_{k,\textrm{real}}$ is its corresponding true angle-delay pair.
Due to space constraints, we only present the normalized RMSE performance
versus SNR as a representative result of the parameter estimation
accuracy.

\subsection{Convergence Behavior}

We now investigate the convergence behavior of the proposed two-timescale
alternating MAP framework. Fig. \ref{fig:overall_convergence} illustrates
the convergence by plotting the channel extrapolation performance
as a function of the iteration number for the outer loop $I_{0}$
at a representative SNR of $10$ dB. It is evident that the proposed
framework converges rapidly, achieving excellent performance within
just a few outer iterations. Consequently, the proposed algorithm
achieves an excellent trade-off between performance and complexity.

\subsection{Influence of SNR}

\begin{figure}
\begin{centering}
\textcolor{blue}{\includegraphics[width=75mm]{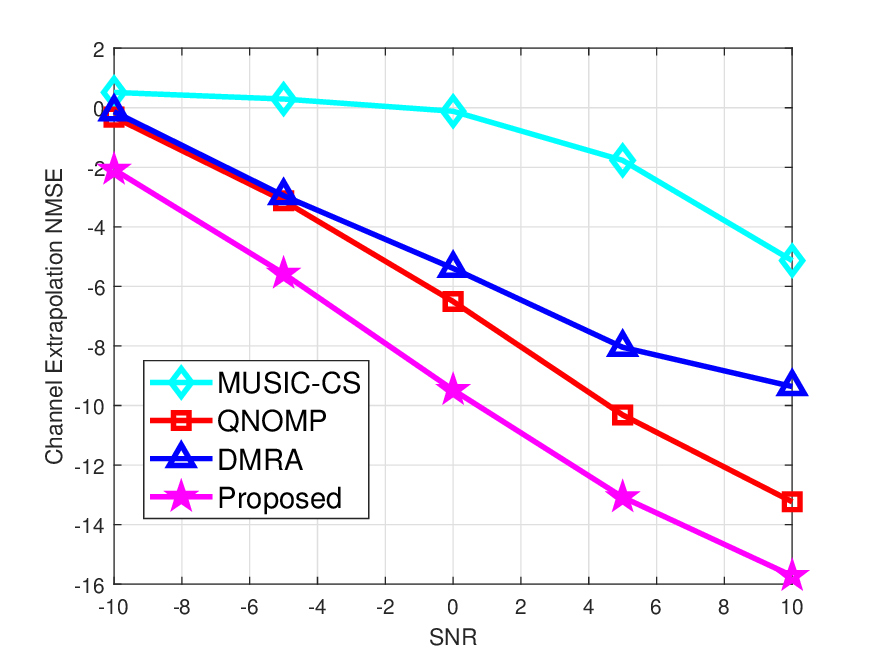}}
\par\end{centering}
\caption{{\small\label{fig:SNR}NMSE of channel extrapolation versus SNR.}}
\end{figure}

In Fig. \ref{fig:SNR}, we compare the NMSE performance of all algorithms
versus SNR, where the number of propagation paths is set to $8$ and
the delay gap between two nearest paths is $30$ ns, which is smaller
than the DFT resolution, i.e., the inverse of one BWP $83.3$ ns.
It can be seen that our proposed two-timescale alternating MAP framework
substantially outperforms all baseline methods across the entire SNR
range, for the following reasons. Firstly, more frequent update of
the active grid enables finer grid refinement and stronger resolution
capability, particularly when dealing with closely spaced paths. In
addition, the proposed tanh-VBI algorithm robustly outputs highly
sparse signals at various SNR levels, facilitating a nearly one-to-one
correspondence between the estimated active grid and the true path
parameters. Thirdly, the BFGS algorithm and step-size carefully designed
by Armijo rule jointly achieve more effective and stable grid update
compared to conventional gradient descent methods. Finally, the uncertain
model parameters such as the noise variance can be automatically learned
based on the VBI framework.

\begin{figure}
\begin{centering}
\textcolor{blue}{\includegraphics[width=75mm]{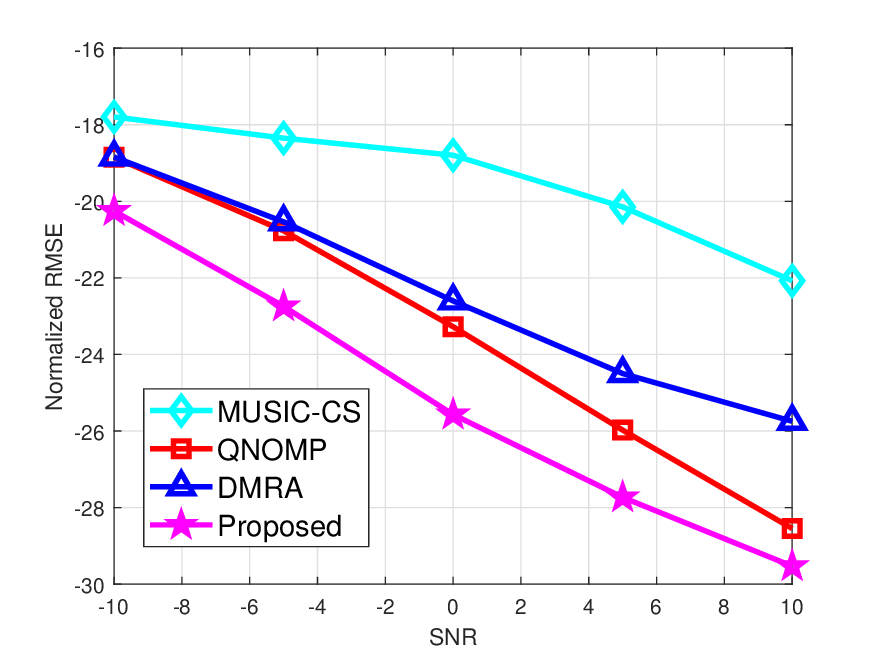}}
\par\end{centering}
\caption{{\small\label{fig:RMSE}Normalized RMSE of parameter estimation versus
SNR.}}
\end{figure}

In Fig. \ref{fig:RMSE}, we compare the normalized RMSE performance
of all algorithms versus SNR with the same configuration. It can be
seen that our proposed two-timescale alternating MAP framework still
outperforms all baseline methods across the entire SNR range, confirming
its superior accuracy in parameter estimation.

It is noteworthy that although the MUSIC-CS algorithm performs well
when the path separation is sufficiently large, its performance dramatically
degrades when the paths are closely spaced, since the the resolution
of the ST-MUSIC algorithm is limited and cannot provide sufficiently
accurate initial grid estimation under such challenging conditions.
The QNOMP algorithm is also worse than the proposed algorithm because
it is more sensitive to the uncertain model parameters such as the
noise variance and the on-gird OMP used in QNOMP is less efficient
than the tanh-VBI. Furthermore, the DMRA algorithm exhibits non-negligible
numerical instability and high sensitivity to hyper-parameters. As
a result, its performance degrades significantly in the high SNR regime.

We further measure the CPU time of each algorithm via MATLAB on a
laptop computer with a 2.5 GHz CPU. For conciseness, we present the
CPU time corresponding to SNR $=5$ dB in Table \ref{tab:CPU_time},
and the results for other SNRs are similar. For a fair comparison,
the number of iterations for each algorithm is set to the minimum
required to achieve convergence. It is observed that the proposed
scheme has a lower runtime than the MUSIC-CS algorithm, while its
runtime is similar with that of the QNOMP and DMRA algorithms. However,
the proposed scheme provides significantly enhanced robustness and
superior estimation performance over all baselines. Considering that
the complexity of our algorithm can be further reduced using inverse-free
methods, as mentioned in \ref{sec:complexity}, the proposed framework
achieves an excellent balance between performance and computational
cost.
\begin{table}[t]
\caption{\label{tab:CPU_time}CPU times of different algorithms.}

\centering{}%
\begin{tabular}{|c|c|}
\hline 
\multicolumn{1}{|c|}{Algorithms} & \multicolumn{1}{c|}{CPU times (s)}\tabularnewline
\hline 
MUSIC-CS & 1.3\tabularnewline
\hline 
QNOMP & 0.35\tabularnewline
\hline 
DMRA & 0.4\tabularnewline
\hline 
The-proposed & 0.45\tabularnewline
\hline 
\end{tabular}
\end{table}

\subsection{Influence of BWP Number}

\begin{figure}
\begin{centering}
\textcolor{blue}{\includegraphics[width=75mm]{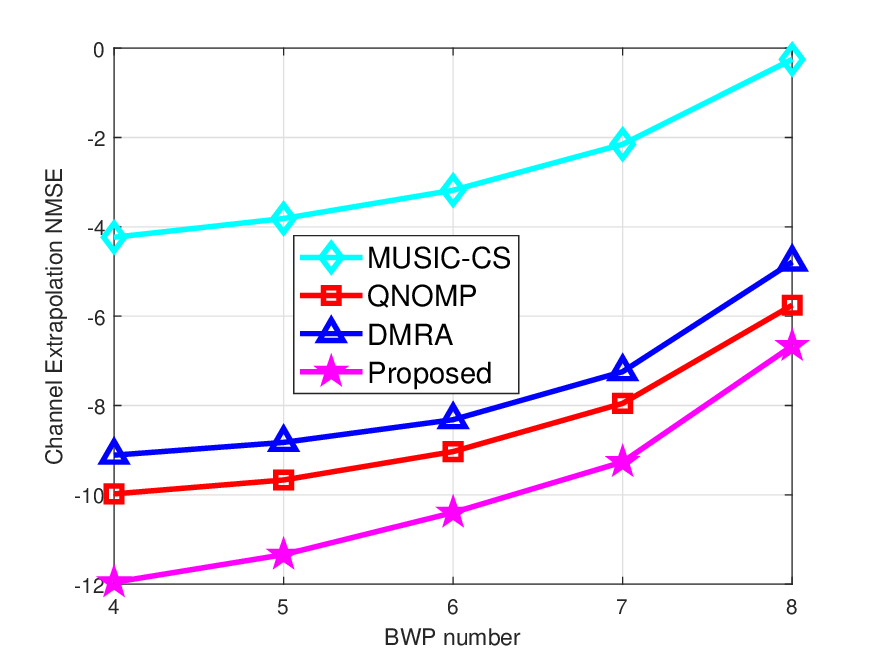}}
\par\end{centering}
\caption{{\small\label{fig:BWP_number}NMSE of channel extrapolation versus
BWP number.}}
\end{figure}

We study the impact of the BWP number on channel extrapolation performance,
where the number of subcarriers in each BWP is fixed at $100$, and
the total BWP number is varied from $4$ to $8$. The number of propagation
paths is set to $8$ and the SNR is set to $5$ dB. As shown in Fig.
\ref{fig:BWP_number}, the channel extrapolation performance of all
algorithms degrades as the BWP number increases, since a larger BWP
number requires higher estimation accuracy of angle and delay parameters.
Nevertheless, the proposed two-timescale alternating MAP framework
consistently achieves the best performance across all tested BWP numbers.
This result demonstrates the robustness and strong super-resolution
capability of our approach, even when the extrapolation task becomes
increasingly challenging.

\subsection{Influence of Antenna Number}

\begin{figure}
\begin{centering}
\textcolor{blue}{\includegraphics[width=75mm]{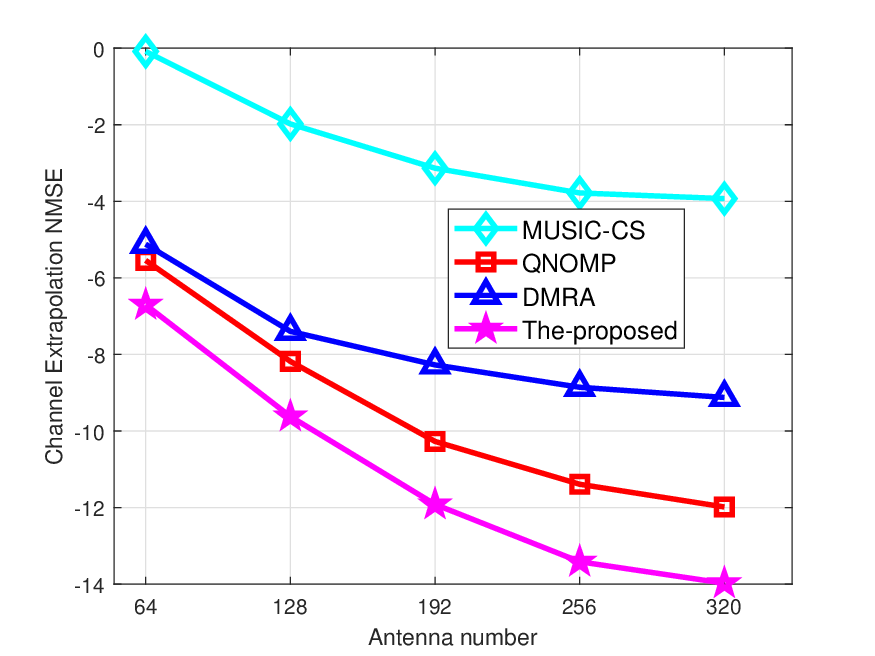}}
\par\end{centering}
\caption{{\small\label{fig:antenna_number}NMSE of channel extrapolation versus
antenna number.}}
\end{figure}

We evaluate the impact of the number of receive antennas on extrapolation
performance. Specially, the total antenna number is varied from $64$
to $320$, where the number of propagation paths is set to $8$ and
the SNR is set to $5$ dB. As shown in Fig. \ref{fig:antenna_number},
increasing the number of antennas improves the channel extrapolation
performance for all algorithms, mainly due to the enhanced spatial
resolution and array gain. Note that due to its sensitivity to hyper-parameters,
the DMRA algorithm exhibits a noticeable performance degradation even
with a large number of antennas. In contrast, our proposed algorithm
consistently delivers the lowest NMSE across all antenna configurations,
further validating its superior robustness and capability for more
accurate parameter estimation and channel extrapolation.

\subsection{Influence of Sparse Prior Model}

\begin{figure}
\begin{centering}
\textcolor{blue}{\includegraphics[width=75mm]{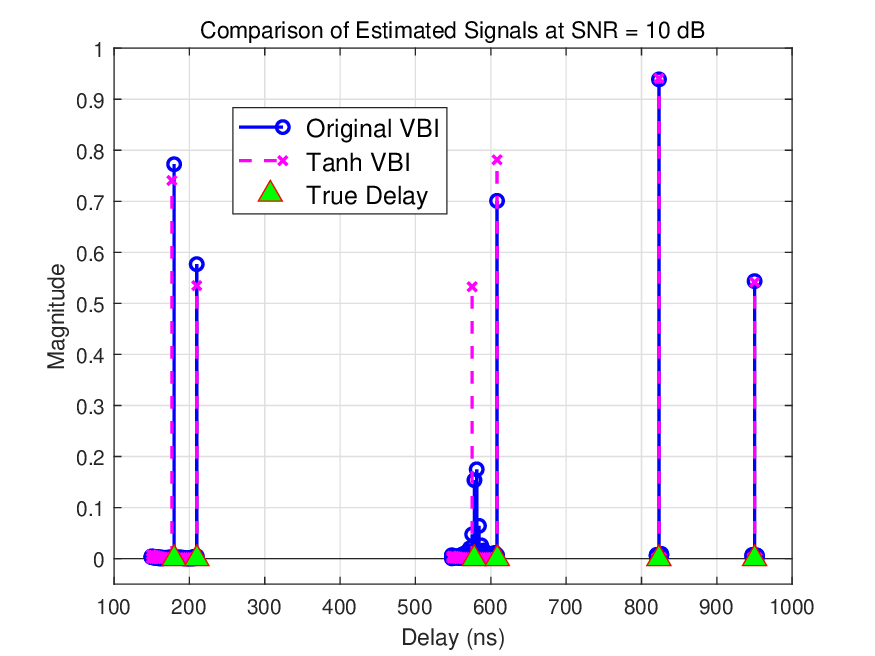}}
\par\end{centering}
\caption{{\small\label{fig:demo_sparsity}An illustration of estimated sparse
signal magnitudes.}}
\end{figure}

We evaluate the impact of the sparse prior model on both the resolution
capability and channel extrapolation performance. To provide a clear
comparison between the original VBI algorithm based on a BGG prior
model and the proposed tanh-VBI algorithm using a BGT prior, we visualize
the estimated sparse signal magnitudes after the first iteration at
SNR = $10$ dB. As shown in Fig. \ref{fig:demo_sparsity}, the proposed
tanh-VBI algorithm is able to robustly produce highly sparse solutions
even under a dense grid, preserving its resolution capability. In
contrast, the original VBI algorithm tends to generate non-sparse
estimation due to its limited sparsity promotion property, which hinders
accurate identification of the active grid elements and effective
super-resolution grid refinement subsequently. 

\begin{figure}
\begin{centering}
\textcolor{blue}{\includegraphics[width=75mm]{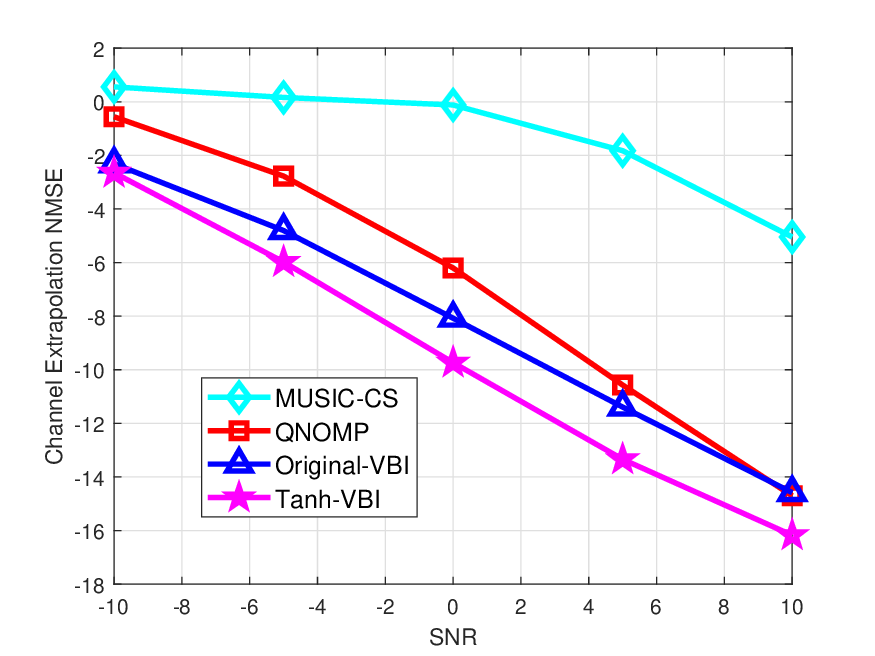}}
\par\end{centering}
\caption{{\small\label{fig:different_VBI}NMSE of channel extrapolation with
different sparse prior models.}}
\end{figure}

In Fig. \ref{fig:different_VBI}, we compare the NMSE performance
of channel extrapolation achieved by the original VBI algorithm with
a BGG sparse prior and the proposed tanh-VBI algorithm with a BGT
sparse prior, where the number of propagation paths is set to $8$.
For a clearer comparison, the performance of the QNOMP and MUSIC-CS
algorithms is also included in the figure. As expected, the proposed
tanh-VBI algorithm demonstrates a significant performance gain over
the original VBI algorithm, especially in the high SNR regime. Note
that although the limited sparsity promotion capability of the original
VBI leads to some performance degradation, it still achieves slightly
better performance than the baseline QNOMP algorithm. These results
highlight the importance of sparse prior model on the resolution capability
and channel extrapolation performance for VBI-based approaches. 

\subsection{Influence of Grid Update Method}

\begin{figure}
\begin{centering}
\textcolor{blue}{\includegraphics[width=75mm]{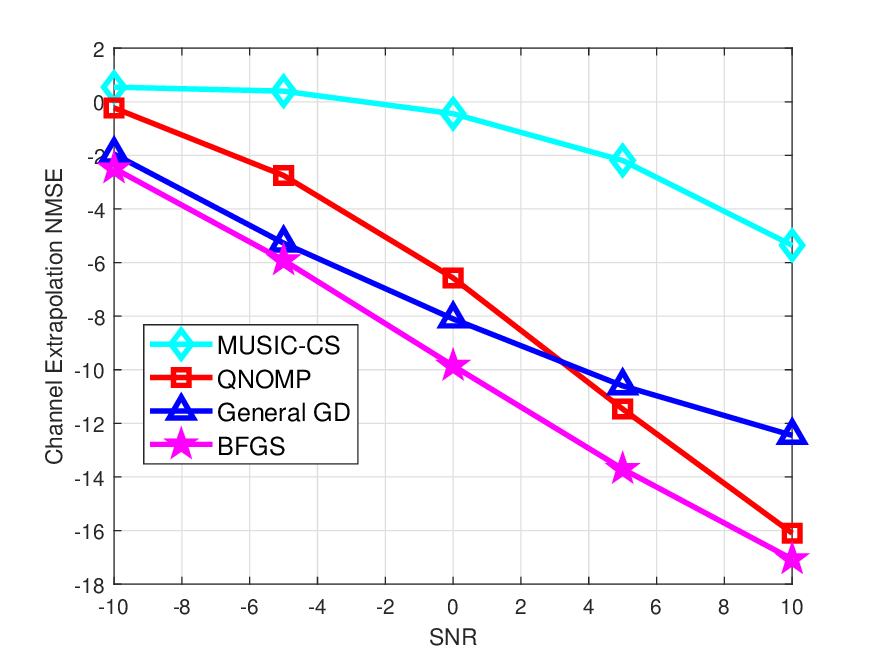}}
\par\end{centering}
\caption{{\small\label{fig:different_grid_update}NMSE of channel extrapolation
with different grid update methods.}}
\end{figure}

We investigate the impact of different grid update methods on channel
extrapolation performance by comparing the conventional gradient descent
method and the BFGS method, where the number of propagation paths
is set to $8$. For a fair comparison, the step-size of both methods
is strictly calculated according to the Armijo rule. As shown in Fig.
\ref{fig:different_grid_update}, the BFGS algorithm consistently
achieves better channel extrapolation performance compared to gradient
descent, with the performance gain being especially significant in
the high SNR regime. The performance improvement is mainly because
the BFGS algorithm can effectively utilize second-order derivative
information of the posterior function, enabling more accurate descent
directions for grid refinement, and is less likely to become trapped
in local optima. 

\section{Conclusion\label{sec:Conlusion}}

\textcolor{black}{In this paper, we proposed a two-timescale alternating
MAP framework for the robust super-resolution compressive sensing
problem. The framework iterates between two key modules operating
at different timescales until convergence: a sparse signal estimation
(SSE) module and a super-resolution grid update (SR-GU) module. First,
for a fixed grid from the SR-GU module, the SSE module leverages a
novel tanh-VBI algorithm on a slow timescale to accurately estimate
the posterior of the sparse signal and identify active grid components
under a dense grid. Subsequently, for a given sparse signal estimate
from the SSE module, the SR-GU module refines the low-dimensional
active grid parameters and their gains on a fast timescale by efficiently
optimizing the likelihood function using the BFGS method. Furthermore,
in the proposed tanh-VBI algorithm, a successive linear approximation
is used to handle the intractable non-linear prior, enabling a closed-form
variational update. Finally, we applied the proposed framework to
the channel extrapolation problem, where simulations showed that our
algorithm achieves significant gains over several state-of-the-art
baseline algorithms.}


\end{document}